\documentclass{SciPost}

\hypersetup{
    colorlinks,
    linkcolor={red!50!black},
    citecolor={blue!50!black},
    urlcolor={blue!80!black}
}

\usepackage[bitstream-charter]{mathdesign}
\usepackage{physics}
\usepackage[capitalise]{cleveref}
\usepackage[dvipsnames, table]{xcolor}
\usepackage{multirow, multicol}
\usepackage{booktabs}
\usepackage{tikz}
\usetikzlibrary{calc, angles, quotes}
\usepackage{subcaption}
\DeclareSymbolFont{usualmathcal}{OMS}{cmsy}{m}{n}
\DeclareSymbolFontAlphabet{\mathcal}{usualmathcal}

\fancypagestyle{SPstyle}{
\fancyhf{}
\lhead{\colorbox{scipostblue}{\bf \color{white} ~SciPost Physics }}
\rhead{{\bf \color{scipostdeepblue} ~Submission }}

\fancyfoot[C]{\textbf{\thepage}}
}

\begin{document}

\pagestyle{SPstyle}

\begin{center}{\Large \textbf{\color{scipostdeepblue}{
%%%%%%%%%% TODO: Write your article's title here
%Article Title, as descriptive as possible, ideally fitting in two lines (approximately 150 characters) or less\\
Spectral optimization of spherical lattices and applications to classical and quantum interacting spin systems
}}}\end{center}

\begin{center}\textbf{
% Use (full) first name (+ middle name initials) + surname format.
% Separate subsequent authors by a comma, omit comma and use "and" for the last author.
% Mark the corresponding author(s) with a superscript symbol in this order
% \star, \dagger, \ddagger, \circ, \S, \P, \parallel, ...
Jonas V\"oller\textsuperscript{1$\star$},
Grigorios Makris\textsuperscript{1},
Fabian Hassler\textsuperscript{2}, and 
Stefan Wessel\textsuperscript{1} 
}\end{center}

\begin{center}
{\bf 1} Institute for Theoretical Solid State Physics, RWTH Aachen University, 52056 Aachen, Germany 
\\
{\bf 2} Institute for Quantum Information, RWTH Aachen University, 52056 Aachen, Germany
%%%%%%%%%% END TODO: AFFILIATIONS
% Provide email address of corresponding author(s)
%\\[\baselineskip]
%$\star$ \href{mailto:email1}{\small jonas.voeller@rwth-aachen.de}\,\quad
%$\dagger$ \href{mailto:email2}{\small email2}
\end{center}

\section*{\color{scipostdeepblue}{Abstract}}
\textbf{\boldmath{%
% The abstract is in boldface, and should fit in 8 lines. It should be written in a clear and accessible style, emphasizing the context, the problem(s) studied, the methods used, the results obtained, the conclusions reached, and the outlook. You can add a table contents, recommended if your paper is more than 6 pages long.
We develop an optimization algorithm for spherical lattice discretizations based on the spectrum of the discretized Laplace operator.
Starting from refined icosahedral lattices, we minimize the splitting of the $(2\ell+1)$, $\ell=0,1,2,\dots$, continuum multiplets expected from $\mathrm{SO}(3)$ symmetry via gradient descent. This approach simultaneously produces  lattices based on triangles with nearly uniform circumradii and perimeters.
Although the optimization uses only the free-particle spectrum, we show that the resulting lattices carry over to interacting models. On the optimized lattices, rotational symmetry is restored for the critical Ising and Potts models and primary scaling dimensions are accurately reproduced for the three-dimensional Ising conformal field theory, when appropriate geometry-dependent interaction strengths are used. Higher-angular-momentum descendant states reveal weak residual symmetry breaking beyond the sectors protected by the underlying icosahedral symmetry.
}}

\vspace{\baselineskip}

%%%%%%%%%% BLOCK: Copyright information
% This block will be filled during the proof stage, and finilized just before publication.
% It exists here only as a placeholder, and should not be modified by authors.
\noindent\textcolor{white!90!black}{%
\fbox{\parbox{0.975\linewidth}{%
\textcolor{white!40!black}{\begin{tabular}{lr}%
  \begin{minipage}{0.6\textwidth}%
    {\small Copyright attribution to authors. \newline
    This work is a submission to SciPost Physics. \newline
    License information to appear upon publication. \newline
    Publication information to appear upon publication.}
  \end{minipage} & \begin{minipage}{0.4\textwidth}
    {\small Received Date \newline Accepted Date \newline Published Date}%
  \end{minipage}
\end{tabular}}
}}
}
%%%%%%%%%% BLOCK: Copyright information

% For convenience during refereeing we turn on line numbers:
% \linenumbers

\vspace{10pt}
\noindent\rule{\textwidth}{1pt}
\tableofcontents
\noindent\rule{\textwidth}{1pt}
\vspace{10pt}

% Write your article contents here, starting from first \section.
% An example structure is given below.
\section{Introduction}
\label{sec:intro}
The non-perturbative numerical formulation of quantum and statistical field theories on curved manifolds remains an important challenge in statistical physics~\cite{2013_Brower, 2021_Brower, 2024_brower_owen}.
One of the most relevant geometries is the sphere, whose continuous rotational symmetry makes it a natural setting for studying conformal field theories (CFTs), which, when mapped from Euclidean $\mathbb{R}^d$ to the cylinder $\mathbb{S}^{d-1}\times \mathbb{R}$, give direct access to operator scaling dimensions via the state-operator correspondence~\cite{1984_Cardy, 1985_Cardy}.
In practice, however, any finite discretization of the sphere necessarily breaks rotational symmetry, and this symmetry breaking introduces systematic errors that become particularly important when extracting continuum properties from lattice simulations~\cite{1987_Alcaraz, 2013_Brower}.

Unlike the Euclidean plane, the sphere cannot be tiled by identical polygons, with the exception given by the five Platonic solids.
In particular, the topology of the sphere requires every triangulation to carry a total topological defect of twelve, distributed among vertices with coordination number unequal to six, preventing the construction of general perfectly regular lattices.
Consequently, spherical lattice constructions must instead seek discretizations that approximate the continuum geometry as closely as possible while minimizing the symmetry breaking introduced by the finite lattice.

Recently, considerable progress has been made in this direction.
Using spherical lattice systems that have been optimized for geometric uniformity in triangle circumradii and perimeter, Brower and collaborators have shown agreement with the $2\mathrm{D}$ Ising CFT in the continuum limit~\cite{2024_brower_owen}.
In one dimension higher, they have studied the $3\mathrm{D}$ Ising CFT using the critical $\phi^4$-theory in the framework of quantum finite elements (QFE), a quantum extension of classical finite elements~\cite{2021_Brower}.
Moreover, an alternative approach to lattice systems is provided by the fuzzy-sphere regularization which relies on a projection of fermions into the lowest spherical Landau level and allows for highly accurate calculations of conformal operator scaling dimensions with remarkably small finite-size effects~\cite{2023_Zhu}.

In this work, we optimize spherical lattices using a physical observable: the spectrum of the discretized Laplace operator.
Since the continuum free particle on $\mathbb{S}^2$ possesses the characteristic $(2\ell+1)$-fold degeneracy pattern dictated by $\mathrm{SO}(3)$ symmetry, the splitting of these multiplets provides a direct quantitative measure of rotational symmetry breaking. The free model serves as a good proxy as obtaining the spectrum is cheap, the degeneracy is exactly known, and directly sensitive to $\mathrm{SO}(3)$ breaking.
We therefore construct a gradient descent optimization algorithm that adjusts the lattice geometry to minimize this spectral splitting.

The central question addressed in this paper is whether optimizing a lattice with respect to the free-particle spectrum also improves the symmetry properties of interacting lattice models, given a model-appropriate coupling construction.
To answer this, we first investigate classical statistical systems on the sphere, including the Ising and Potts models at criticality.
We then study the quantum transverse-field Ising model (TFIM), whose quantum critical point belongs to the $3\mathrm{D}$ Ising universality class via the quantum-to-classical mapping~\cite{1976_Suzuki}, and which, in the continuum, provides access to the CFT on $\mathbb{S}^{2}\times\mathbb{R}$, where radial quantization relates excitation energies to scaling dimensions of CFT operators~\cite{1985_Cardy}.

Our results show that spectral optimization substantially improves the rotational symmetry of the lattice Laplacian and simultaneously drives the lattice towards the geometric uniformity conditions underlying previously proposed isoradial and isoperimetric constructions~\cite{2024_brower_owen}.
As a consequence, the optimized lattices recover rotational symmetry in the critical Ising and Potts models.
For the TFIM, we accurately determine the critical transverse field and the critical Binder cumulant.
The extracted ratio of primary scaling dimensions agrees well with bootstrap predictions~\cite{2015_simmonsduffin, 2014_simmonsduffin, 2014_El_Showk, 2012_El_Showk}, demonstrating that the optimized lattices accurately capture the lowest sectors of the $3\mathrm{D}$ Ising CFT.
However, an analysis of descendant operator scaling dimensions reveals weak residual symmetry breaking for states with $\ell>2$ that are not symmetry protected by the underlying icosahedral structure.
This indicates that, although the optimized lattices combined with appropriate couplings provide a substantial improvement over standard icosahedral discretizations, lattice artifacts remain visible in higher-angular-momentum excitations in the quantum case.

The paper is organized as follows.
In \cref{sec:latticeoptimization}, we introduce the spherical lattice construction based on an icosahedral seed and the corresponding spectral optimization algorithm.
We analyze the free-particle spectrum and demonstrate the restoration of rotational symmetry as quantified by the deviation from the continuum degeneracy pattern.
In \cref{sec:isingmodelons2} and \cref{sec:pottsmodelons2}, we utilize the optimized lattices paired with suitable couplings from the isoradial graph construction to investigate the critical Ising and Potts models and test rotational invariance through spherical-harmonic decompositions of spin-spin correlation functions.
Finally, in \cref{sec:tfimons2}, we study the TFIM  at its quantum critical point, determined by finite-size scaling based on Binder cumulants, and examine the state-operator correspondence of primary and descendant CFT operators.

\section{Lattice Optimization via Laplacian gradient descent}
\label{sec:latticeoptimization}
We begin by constructing an initial spherical lattice.
A natural starting point is the regular icosahedron, consisting of twelve vertices and $20$ faces, each of which is an equilateral triangle.
As one of the five Platonic solids, the regular icosahedron provides a highly symmetric geometry in which all vertices and faces are equivalent.
Moreover, its rotational symmetry group  $I\simeq A_5$ is of order $60$, the largest among the rotational symmetry groups of the Platonic solids~\cite{1973_Coxeter}.

\begin{figure}[h!]
    \centering
    \includegraphics[width=0.8\textwidth]{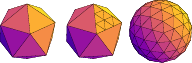}
    \caption{Initialization of a spherical lattice starting from an icosahedral seed with a subdivision level of $k=3$. (figure adapted from~\cite{2024_brower_owen})}\label{fig:sphereinitialization}
\end{figure}

Following~\cite{2024_brower_owen}, we refine this discretization by subdividing each triangular face into smaller equilateral triangles, dividing each edge into $k$ equal segments.
New vertices are introduced at the resulting intersections and subsequently rescaled radially to lie on the unit sphere (\cref{fig:sphereinitialization}).
This subdivision preserves the twelve original vertices as five-fold coordinated defects, while the newly introduced vertices are six-fold coordinated.
The resulting $k$-th subdivision yields a lattice containing $N=2+10k^2$ vertices~\cite{2024_brower_owen}.

\subsection{Free-particle model on \texorpdfstring{$\mathbb{S}^2$}{S2}}
We now introduce the free-particle model on our discretized lattices.
Due to its analytical solvability and clear connection to spherical symmetry via the Laplace operator, we anticipate that a related optimization algorithm will produce geometries that exhibit enhanced $\mathrm{SO}(3)$ symmetry for the free-particle model itself and for other models, given a suitable adjustment of the couplings.

In the continuum, the free particle on the sphere has the well-known energy spectrum $E_\ell=\ell(\ell+1)/2$, $\ell=0,1,2,\dots$, with each energy level exhibiting a $(2\ell+1)$-fold degeneracy.
This degeneracy follows from the $\mathrm{SO}(3)$ symmetry of the Laplace operator in the Hamiltonian $H=-\nabla^2/2$ on the sphere, where we set $\hbar$ and the particle mass to unity.
On a discretized lattice, the continuous rotational symmetry is reduced to the finite symmetry group of the lattice, which generally lifts the continuum degeneracies.
We therefore use the free-particle spectrum as a diagnostic of residual spherical symmetry and subsequently optimize the lattice geometry with respect to the resulting spectral deviations.

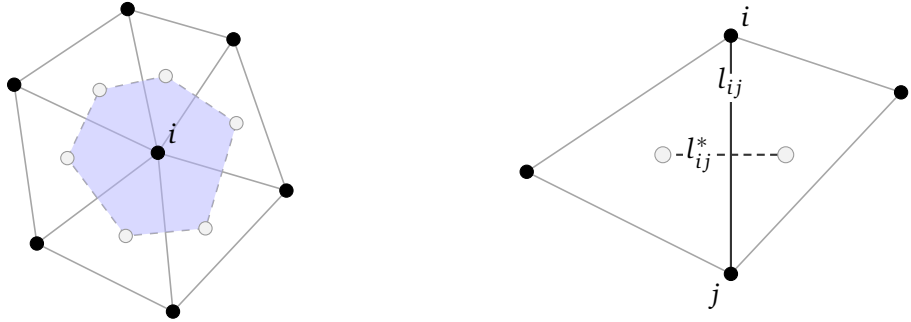
\begin{figure}[h!]
    \centering
    \begin{subfigure}[t]{0.49\textwidth}
        \centering
        \begin{tikzpicture}[scale=1]
\coordinate (A) at (0,0.9);
\coordinate (B) at (1.8, 0);
\coordinate (C) at (1.6, 2.1);
\coordinate (D) at (3.3, 1.6);
\coordinate (E) at (-0.3,3.0);
\coordinate (F) at (1.2, 4.0);
\coordinate (G) at (2.6, 3.6);

\coordinate (CC1) at (1.175, 1.0);
\coordinate (CC2) at (0.4032, 2.029);
\coordinate (CC3) at (0.83, 2.93);
\coordinate (CC4) at (1.704, 3.114);
\coordinate (CC5) at (2.6385, 2.491);
\coordinate (CC6) at (2.2295, 1.1004);

%\draw[color=blue!50, thin, densely dotted](CC1) circle (1.17925);

\draw[gray!70, semithick] (C) -- (A);
\draw[gray!70, semithick] (C) -- (B);
\draw[gray!70, semithick] (C) -- (D);
\draw[gray!70, semithick] (C) -- (E);
\draw[gray!70, semithick] (C) -- (F);
\draw[gray!70, semithick] (C) -- (G);

\draw[gray!70, semithick] (A) -- (B);
\draw[gray!70, semithick] (A) -- (E);

\draw[gray!70, semithick] (B) -- (D);
\draw[gray!70, semithick] (D) -- (G);
\draw[gray!70, semithick] (F) -- (G);
\draw[gray!70, semithick] (E) -- (F);

\draw[gray!70, semithick, dashed] (CC1) -- (CC2);
\draw[gray!70, semithick, dashed] (CC2) -- (CC3);
\draw[gray!70, semithick, dashed] (CC3) -- (CC4);
\draw[gray!70, semithick, dashed] (CC4) -- (CC5);
\draw[gray!70, semithick, dashed] (CC5) -- (CC6);
\draw[gray!70, semithick, dashed] (CC6) -- (CC1);

\fill[blue!30, opacity=0.5] (CC1) -- (CC2) -- (CC3) -- (CC4) -- (CC5) -- (CC6) -- cycle;

\filldraw[black] (A) circle (2.5pt);
\filldraw[black] (B) circle (2.5pt);
\filldraw[black] (C) circle (2.5pt) node[anchor=south west]{$i$};
\filldraw[black] (D) circle (2.5pt);
\filldraw[black] (E) circle (2.5pt);
\filldraw[black] (F) circle (2.5pt);
\filldraw[black] (G) circle (2.5pt);

\filldraw[solid, color=gray!80, fill=gray!10, thin] (CC1) circle (2.5pt);
\filldraw[solid, color=gray!80, fill=gray!10, thin] (CC2) circle (2.5pt);
\filldraw[solid, color=gray!80, fill=gray!10, thin] (CC3) circle (2.5pt);
\filldraw[solid, color=gray!80, fill=gray!10, thin] (CC4) circle (2.5pt);
\filldraw[solid, color=gray!80, fill=gray!10, thin] (CC5) circle (2.5pt);
\filldraw[solid, color=gray!80, fill=gray!10, thin] (CC6) circle (2.5pt);
\end{tikzpicture}
    \end{subfigure}
    \begin{subfigure}[t]{0.49\textwidth}
        \centering
        \begin{tikzpicture}[scale=1.5]
\coordinate (A) at (0,0.9);
\coordinate (B) at (1.8, 0);
\coordinate (C) at (1.8, 2.1);
\coordinate (D) at (3.3, 1.6);
\coordinate (CC1) at (1.2, 1.05);
\coordinate (CC2) at (2.28333, 1.05);

\draw[gray!70, semithick] (A) -- (B);
\draw[gray!70, semithick] (A) -- (C);
\draw[gray!70, semithick] (B) -- (D);
\draw[gray!70, semithick] (C) -- (D);

\draw[black!80, thick] (B) -- ($(B)!0.76!(C)$);
\node[fill=white, inner sep=1pt] at ($(B)!0.8!(C)$) {$l_{ij}$};
\draw[black!80, thick] ($(B)!0.84!(C)$) -- (C);

\draw[densely dashed, black!80, thick] (CC1) -- ($(CC1)!0.26!(CC2)$);
\node[fill=white, inner sep=1pt] at ($(CC1)!0.3!(CC2)$) {$l^*_{ij}$};
\draw[densely dashed, black!80, thick] ($(CC1)!0.44!(CC2)$) -- (CC2);

\filldraw[black] (A) circle (1.666pt);
\filldraw[black] (B) circle (1.666pt) node[anchor=north east]{$j$};
\filldraw[black] (C) circle (1.666pt) node[anchor=south west]{$i$};
\filldraw[black] (D) circle (1.666pt);
\filldraw[solid, color=gray!80, fill=gray!10, thin] (CC1) circle (2pt); %node[anchor=north east, color=black]{$i^*$};
\filldraw[solid, color=gray!80, fill=gray!10, thin] (CC2) circle (2pt); %node[anchor=north east, color=black]{$j^*$};
\end{tikzpicture}
    \end{subfigure}
    \caption{Construction of the Laplacian couplings. Left: the dual lattice, obtained by connecting the circumcenters (open circles) of neighboring triangles, encloses the Voronoi cell of vertex $i$ (shaded blue). Right: the coupling $J_{ij}$ is set by half the ratio of the shared dual edge length $l^*_{ij}$ to the direct lattice edge length $l_{ij}$.}\label{fig:duallatticehoppings}
\end{figure}

To discretize the Laplacian, we associate each vertex $i$ with a position $\hat{\mathbf{n}}_i$ on the unit sphere and introduce the tight-binding Hamiltonian
\begin{equation}
    H_\text{lat}=\sum_i \mu_i \ket{i}\bra{i} -\sum_{\langle ij \rangle}J_{ij} \ket{i}\bra{j},
\end{equation}
where the sum over $\langle ij \rangle$ runs over neighboring vertices.
The lattice Hamiltonian approaches the continuum Laplacian in the refinement limit when the on-site terms are chosen as $\mu_i=\sum_j J_{ij}$ and the hopping coefficients are given by~\cite{1982_Christ_Friedberg_Lee}
\begin{equation}\label{eq:laplaciancouplings}
    J_{ij}=J_{ji}=\frac{l_{ij}^*}{2l_{ij}},
\end{equation}
where $l_{ij}=\lvert\hat{\mathbf{n}}_i-\hat{\mathbf{n}}_j\rvert$ denotes the edge length between neighboring sites $i$ and $j$, and $l_{ij}^*$ is the corresponding length of the dual edge separating their Voronoi cells (\cref{fig:duallatticehoppings}).

The lattice geometry is treated as a piecewise-flat triangulated surface.
Accordingly, the edge lengths and associated geometric quantities are defined within the piecewise-flat lattice geometry rather than by geodesic distances on $\mathbb{S}^2$, following the framework of Regge calculus~\cite{1961_Regge}.

\begin{figure}[h!]
    \centering
    \includegraphics[width=0.9\textwidth]{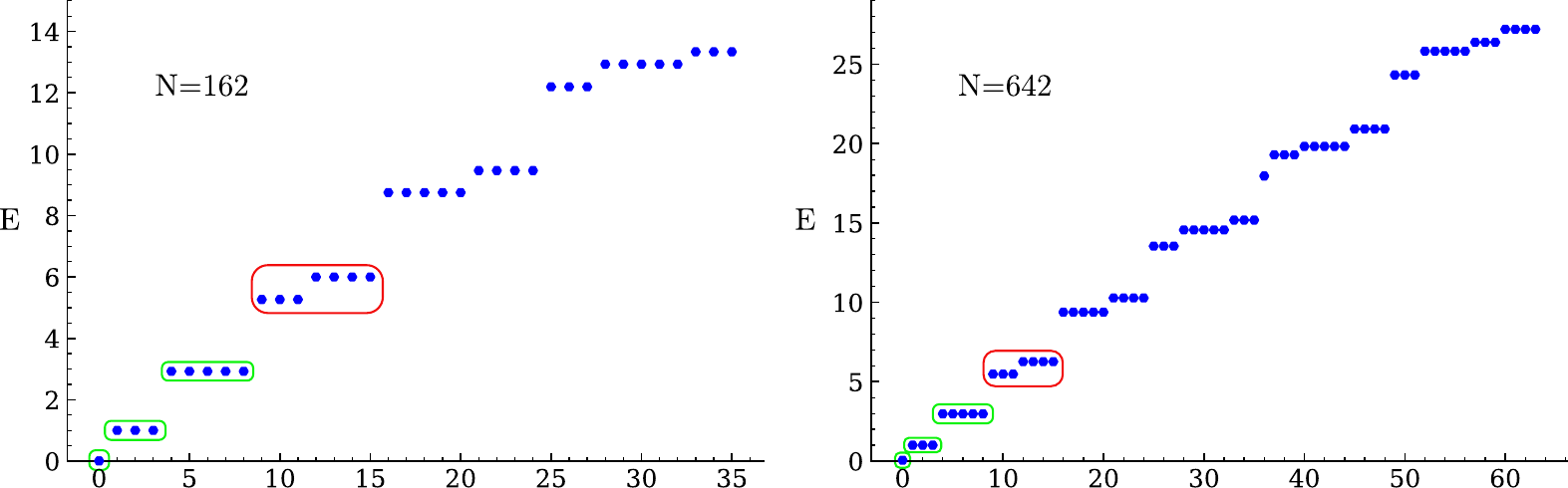}
    \caption{Energy spectrum of the free-particle Hamiltonian for base icosahedral lattices with $N=162$ (left) and $N=642$ (right) vertices.
    Both spectra exhibit the expected continuum degeneracy only for $\ell\le2$, which we indicate with green outlines.
    We furthermore indicate the first level with degeneracy splitting, corresponding to $\ell=3$, with red outlines. 
    The labeling of the x-axis indicates the indexing of the energy levels.}\label{fig:basespectra}
\end{figure}

Diagonalizing $H_\text{lat}$ for the base icosahedral lattices with $N=162$ and $N=642$ vertices yields the spectra shown in \cref{fig:basespectra}.
For comparison with the continuum spectrum, we rescale the energy levels so that the first excited level satisfies $E_1=1$.

The levels corresponding to $\ell\le2$ retain their full continuum degeneracies.
This is consistent with the symmetry reduction from the full spherical symmetry to the icosahedral symmetry group $\mathrm{O}(3)\to I_h$.
The continuum representations associated with $\ell=0,1,2$ remain irreducible when restricted to $I_h$, corresponding to the irreducible representations (irreps) $A_g$, $T_{1u}$, and $H_g$, respectively (details on the group theoretical background are given in Appendix~\ref{sec:grouptheorydetails}).
Consequently, the corresponding lattice energy levels remain degenerate.

For $\ell\ge3$, the continuum representations generally decompose into direct sums of multiple irreps of $I_h$, leading to a splitting of the corresponding energy levels.
For example, the seven-dimensional $\ell=3$ representation decomposes according to $\ell=3\to T_{2u}\oplus G_u$, with dimensions $\text{dim}(T_{2u})=3$ and $\text{dim}(G_u)=4$.
The corresponding sevenfold-degenerate continuum level therefore splits into two lattice levels with degeneracies three and four.
The observed splitting of the higher-energy levels is thus a direct manifestation of the reduced symmetry of the base icosahedral lattices, and it provides a diagnostic of deviations from spherical symmetry that serves as a basis for optimizing the lattice geometry in the remainder of this section.

\subsection{Gradient descent optimization algorithm}
To reduce the symmetry breaking introduced by the discretization, we propose an optimization algorithm that adjusts the positions of the vertices on the unit sphere by gradient descent on a loss function that penalizes energy splittings within each continuum-degenerate level.

Let $\mathcal{D}$ denote the set of continuum energy multiplets included in the optimization, where each multiplet labeled by $\ell$ contains $2\ell+1$ states.
Excluding the trivially non-degenerate $\ell=0$, we define the loss function as 
\begin{equation}\label{eq:lossfunction}
    \mathcal{L}_\text{deg} = \sum_{\ell \in \mathcal{D}} \frac{1}{(2\ell+1)^2} \sum_{n=1}^{2\ell+1} {\left(E_n - \bar{E}_\ell\right)}^2,
\end{equation}
where $E_n$ are the eigenvalues of $H_{\text{lat}}$ and $\bar{E}_\ell$ is the mean energy of level $\ell$.
This way, the loss function sums up the variances across each $\ell$ multiplet, weighted by an additional factor $1/(2\ell+1)^2$  to prioritize low-lying energy states, which was found to increase numerical stability.
Its gradient with respect to a single vertex at position $\hat{\textbf{n}}_k$ is given by 
\begin{equation}\label{eq:gradient}
    \nabla_{\hat{\textbf{n}}_k} \mathcal{L}_{\text{deg}} = \sum_{\ell\in \mathcal{D}} \frac{2}{(2\ell+1)^2} \sum_{n=1}^{2\ell+1} \left(E_n-\bar{E}_\ell\right)
    \sum_{ij} \left( \frac{\partial E_n}{\partial J_{ij}} -\frac{1}{2\ell+1} \sum_{m=1}^{2\ell+1} \frac{\partial E_m}{\partial J_{ij}}\right) \nabla_{\hat{\textbf{n}}_k} J_{ij}.
\end{equation}
Evaluating this gradient requires the partial derivatives of the eigenvalues with respect to the hopping coefficients and the gradient of those hopping coefficients with respect to the vertex position.
For the former, we know that the variation of a single hopping term $J_{ij}$ results in changes in both the diagonal and off-diagonal Hamiltonian matrix entries according to
\begin{equation}
    \frac{\partial H_\text{lat}}{\partial J_{ij}} = \ket{i}\bra{i} + \ket{j}\bra{j} - \ket{i}\bra{j} - \ket{j}\bra{i}.
\end{equation}
Thus, applying first-order perturbation theory yields
\begin{equation}
    \frac{\partial E_n}{\partial J_{ij}} = \Big\langle\psi_n\Big\rvert\frac{\partial H_\text{lat}}{\partial J_{ij}}\Big\lvert\psi_n\Big\rangle =  (\psi_n(i)-\psi_n(j))^2,
\end{equation}
where $\psi_n(i)$ denotes the entry $i$ of the eigenvector $n$.
In isolation, this quantity is only correct for non-degenerate subspaces, since in the presence of degeneracies, individual eigenvalue derivatives are basis-dependent.
However, since all states within a degenerate subspace enter the loss with the same coefficient, only their sum is required, which is the basis-independent trace of the perturbation projected onto the degenerate subspace.

For the remaining gradient $\nabla_{\hat{\textbf{n}}_k} J_{ij}$, a variation in the position of a single vertex only affects hopping coefficients that depend on that vertex and $2n_c$ surrounding vertices, where $n_c$ is the coordination number of that vertex. 
We thus find an analytical form of $\nabla_{\hat{\textbf{n}}_k}J_{ij}$, so that the geometric part of the gradient evaluation scales only linearly with the local coordination number (for a full derivation, cf. Appendix~\ref{sec:gradientderivation}).

\subsubsection{Optimization effects on energy spectra}
\begin{figure}[h!]
    \centering
    \includegraphics[width=0.9\textwidth]{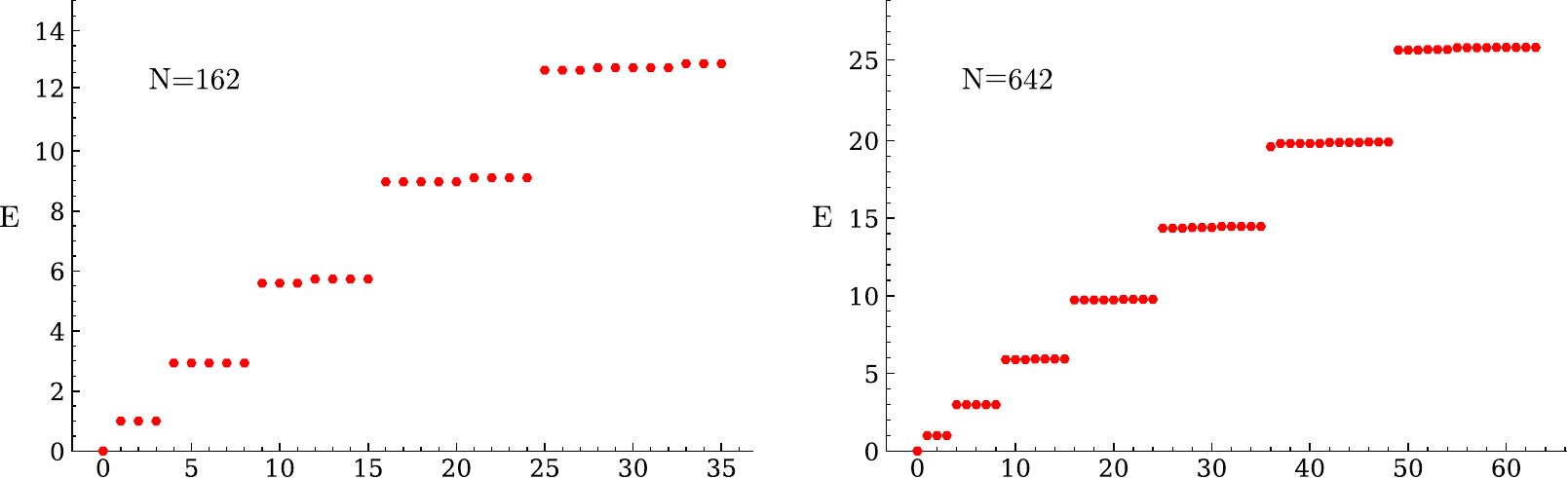}
    \caption{Energy spectrum of the free-particle Hamiltonian for optimized icosahedral lattices with $N=162$ (left) and $N=642$ (right) vertices.
    The degeneracy of levels with $\ell\le2$ is still exact.
    Furthermore, the optimized lattices exhibit substantially improved degeneracy for higher angular momentum energy levels.
    The labeling of the x-axis again indicates the indexing of the energy levels.}\label{fig:modspectra}
\end{figure}

Minimizing \cref{eq:lossfunction} by employing \cref{eq:gradient} on the base icosahedral lattices yields geometries with substantially reduced level splitting, as shown in \cref{fig:modspectra}.
Across the refinements studied, we optimize with respect to twelve energy levels for most lattices (more detail on the number of multiplets used can be found in Appendix~\ref{sec:laplacianappendix}), leading to a decrease in $\mathcal{L}_\text{deg}$ by roughly one to three orders of magnitude relative to the unoptimized lattice, with the largest gains at the finest resolutions\footnote{Note that larger improvements are achievable using this method, but we stop the optimization according to a uniformity-based criterion introduced in \cref{sec:geometriceffects}}.
We verified that the optimized vertices retain the original triangulation and that the resulting displacements vary smoothly across the lattice.
In the refinement limit, we expect the spectra of the optimized lattices to converge to the exact analytical result, indicating geometries with substantially improved spherical symmetry.

\subsubsection{Optimization effects on geometric uniformity}\label{sec:geometriceffects}
\begin{figure}[h!]
    \centering
    \begin{subfigure}[t]{0.45\textwidth}
        \centering
        \includegraphics[width=0.8\textwidth]{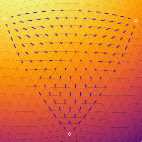}
    \end{subfigure}
    \begin{subfigure}[t]{0.45\textwidth}
        \centering
        \includegraphics[width=0.8\textwidth]{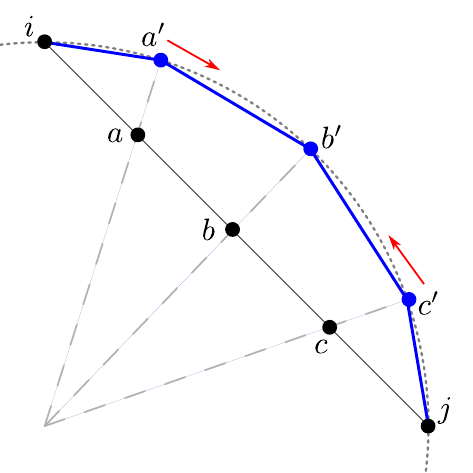}
    \end{subfigure}
    \caption{Movement of the vertices during optimization.
    Left: the background shows the base icosahedral lattice.
    Blue arrows indicate where vertices will end up after optimizing.
    The arrows are only shown for one of the original 20 faces.
    The three vertices of the original regular icosahedron enclosing that face are indicated with white open circles.
    Right: sketch of how the optimization evens out the non-uniform projection of the initial lattice construction.}\label{fig:vertexmovement}
\end{figure}

Comparing the geometry of the lattices before and after optimization (\cref{fig:vertexmovement}), we observe that the optimized vertex displacements are directed toward the centers of the original icosahedral faces.
This movement has the effect of compensating for the prior non-uniform radial projection, which stretches edges near the center of each original icosahedral face relative to those near its corners.
The gradient vanishes at the twelve original icosahedron vertices due to the (fivefold) symmetry around each vertex.

Thus, although the optimization is defined entirely in terms of the spectral properties of the discretized Laplacian, it also has a direct effect on the geometric uniformity of the lattice.
To quantify this effect, we measure the relative spread in circumradius $R$ and perimeter $P$ across all triangles in the lattices, defined as~\cite{2024_brower_owen}
\begin{equation}
    \nu_R=\frac{\langle R^2\rangle}{\langle R\rangle^2}-1 \quad \text{and} \quad \nu_P=\frac{\langle P^2\rangle}{\langle P\rangle^2}-1.
\end{equation}
These uniformity measures are known to be important in the definition of critical couplings for interacting theories, including the Ising and Potts models~\cite{2010_Boutillier, 2018_Duminil_Copin, 2024_brower_owen}, cf. the following sections. 

During the optimization process, both the circumradius and perimeter variances decrease together with the spectral loss, but reach their minima slightly before the spectral loss itself (for corresponding loss plots and additional details on the gradient descent optimization, cf. Appendix~\ref{sec:laplacianappendix}).
Motivated by the importance of the geometric uniformity for the critical couplings used below, we choose the minimum of the combined uniformity measures as our stopping criterion, rather than minimizing the spectral loss further.

\begin{figure}[]
    \centering
    \includegraphics[width=0.5\textwidth]{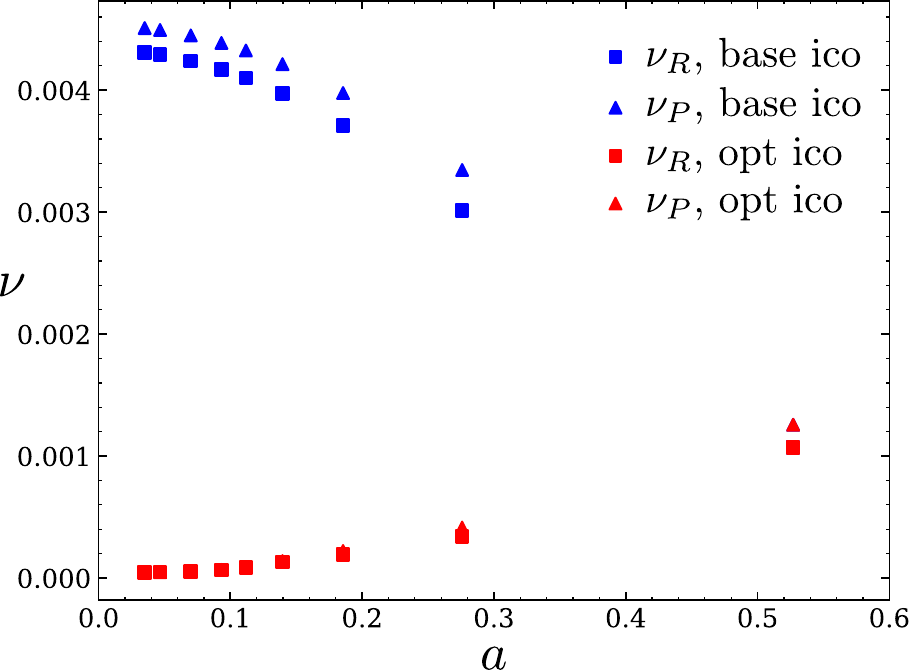}
    \caption{Measurement of uniformity in the circumradii (squares) and perimeters (triangles) as a function of the lattice spacing on the base (blue) and optimized (red) icosahedral lattices.}\label{fig:uniformity}
\end{figure}

The uniformity of the base and resulting optimized lattices with respect to the lattice spacing $a$ is shown in \cref{fig:uniformity}.
Here, we characterize the lattice spacing by $a=\langle A\rangle^{1/2}$, where $\langle A\rangle$ denotes the average Voronoi cell area~\cite{2024_brower_owen}.
The base icosahedral lattices exhibit a nonzero circumradius and perimeter non-uniformity that remains finite as $a\to0$, while both $\nu_R$ and $\nu_P$ decrease towards zero for the optimized lattices in the refinement limit.

In their work, Brower and Owen explicitly optimize the vertex positions with respect to the uniformity conditions imposed by their Ising coupling construction~\cite{2024_brower_owen}, which are exactly the uniformity in circumradii and perimeter.
Interestingly, their explicitly geometric optimization and our spectral optimization appear to drive the lattices to a similarly geometrically uniform regime.
This suggests that the spectral degeneracy pattern is sensitive to the same geometric non-uniformities that enter the construction of the critical couplings.

% !TeX root = ../SciPostPhys.tex

\section{Ising model on \texorpdfstring{$\mathbb{S}^2$}{S2}}
\label{sec:isingmodelons2}
Having established that spectral optimization substantially reduces rotational symmetry breaking in the discretized Laplacian and drives the lattices towards geometric uniformity, we now investigate whether these improvements extend to an interacting critical theory.
As a first test, we consider the Ising model on the sphere with Hamiltonian
\begin{equation}
H=-\sum_{\langle ij \rangle} J_{ij} \sigma_i \sigma_j, \; \sigma_i=\pm1,
\end{equation}
so that each Ising model configuration is weighted by the Boltzmann factor $\exp(-\beta H)$ at an inverse temperature $\beta=1/T$.
For this model, critical couplings can be constructed exactly on isoradial graphs~\cite{2010_Boutillier}.
The construction requires every face of the lattice to be inscribed in a circle of uniform radius and that the corresponding circumcenter lies within the face.
The latter condition is automatically satisfied for the lattices considered here, while the former is asymptotically realized by the optimized lattices, as shown in the previous section.

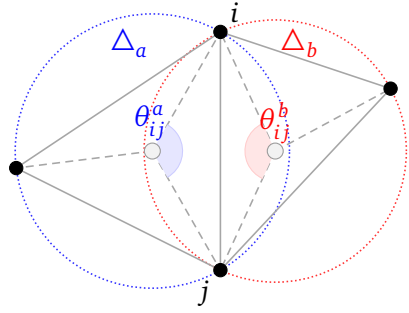
\begin{figure}[]
    \centering
    \begin{tikzpicture}[scale=1.5]
\coordinate (A) at (0,0.9);
\coordinate (B) at (1.8, 0);
\coordinate (C) at (1.8, 2.1);
\coordinate (D) at (3.3, 1.6);
\coordinate (CC1) at (1.2, 1.05);
\coordinate (CC2) at (2.28333, 1.05);

\draw[color=blue!80, semithick, densely dotted](CC1) circle (1.209);
\draw[color=red!80, semithick, densely dotted](CC2) circle (1.156);

\draw[gray!70, semithick] (A) -- (B);
\draw[gray!70, semithick] (A) -- (C);
\draw[gray!70, semithick] (B) -- (C);
\draw[gray!70, semithick] (B) -- (D);
\draw[gray!70, semithick] (C) -- (D);

\draw[gray!70, semithick, densely dashed] (CC1) -- (A);
\draw[gray!70, semithick, densely dashed] (CC1) -- (B);
\draw[gray!70, semithick, densely dashed] (CC1) -- (C);
\draw[gray!70, semithick, densely dashed] (CC2) -- (B);
\draw[gray!70, semithick, densely dashed] (CC2) -- (C);
\draw[gray!70, semithick, densely dashed] (CC2) -- (D);

\draw pic [draw=blue!20, fill=blue!10, angle radius=0.4cm] {angle = B--CC1--C};\node[text=blue] at (1.18, 1.3) {$\theta_{ij}^a$};
\draw pic [draw=red!20, fill=red!10, angle radius=0.4cm] {angle = C--CC2--B};\node[text=red] at (2.28, 1.3) {$\theta_{ij}^b$};

\node[text=blue] at (1.0, 2.0) {$\triangle_a$};
\node[text=red] at (2.5, 2.0) {$\triangle_b$};

\filldraw[black] (A) circle (1.666pt);
\filldraw[black] (B) circle (1.666pt) node[anchor=north east]{$j$};
\filldraw[black] (C) circle (1.666pt) node[anchor=south west]{$i$};
\filldraw[black] (D) circle (1.666pt);
\filldraw[solid, color=gray!80, fill=gray!10, thin] (CC1) circle (2pt); %node[anchor=north east, color=black]{$i^*$};
\filldraw[solid, color=gray!80, fill=gray!10, thin] (CC2) circle (2pt); %node[anchor=north east, color=black]{$j^*$};
\end{tikzpicture}
    \caption{Rhombus construction from two cells of a triangular lattice that
    defines the coupling $J_{ij}$.
    In a truly isoradial graph, $\theta^a_{ij}=\theta^b_{ij}$, so that the rhombus enclosing the edge $\langle ij\rangle$ is uniquely defined.}\label{fig:isoradial}
\end{figure}

For a triangular lattice with globally uniform circumradius, the critical coupling on edge $\langle ij\rangle$ is fixed by the local rhombus angle $\theta_{ij}=2\arcsin(l_{ij}/(2R_{ijk}))$ subtended at the circumcenter of the adjacent triangle (\cref{fig:isoradial}), where $R_{ijk}=l_{ij}l_{jk}l_{ki}/(4A^\triangle_{ijk})$ is the circumradius of the triangle with area $A^\triangle_{ijk}$~\cite{2018_Duminil_Copin}.
The corresponding critical coupling is~\cite{2010_Boutillier}\footnote{We use the angle convention defined in~\cite{2018_Duminil_Copin}, which differs from the rhombus angles $\theta_e$ used in~\cite{2010_Boutillier} by $\theta_e=(\pi-\theta_{ij})/2$.}
\begin{equation}\label{eq:isingcouplings}
    K_{ij}=\beta J_{ij}=\frac12\log\left(\frac{1+\cos(\theta_{ij}/2)}{\sin(\theta_{ij}/2)}\right).
\end{equation}

At finite lattice spacing, the circumradii of the two triangles sharing an edge are generally not identical, and the two adjacent triangles yield slightly different values of the coupling in \cref{eq:isingcouplings}.
We therefore assign a unique coupling to each edge by averaging the two values, $K_{ij}=(K_{ij}^{\triangle_a}+K_{ij}^{\triangle_b})/2$.
In the continuum limit, the two definitions coincide as the circumradius becomes globally uniform.

As an independent check of this construction, we compare the resulting couplings with those of Brower and Owen~\cite{2024_brower_owen}, who derive critical Ising couplings using Kramers-Wannier duality and a Wilson-Majorana fermion loop expansion.
Despite the different derivations, we find that on both the base and optimized icosahedral lattices, the two coupling definitions agree to machine precision ($\lesssim10^{-15}$ relative error), providing a nontrivial cross-check between two independently derived routes to the critical point.

\subsection{Rotational symmetry of the Ising model on \texorpdfstring{$\mathbb{S}^2$}{S2}}
\label{subsec:rotsymmisingons2}
We now test directly whether the optimized lattices restore rotational symmetry in the interacting Ising model.
Following~\cite{2024_brower_owen}, we analyze the spin-spin correlation function by expanding it in the basis of spherical harmonics,
\begin{equation}\label{eq:rotsymmobservable}
    \langle\sigma(\hat{\mathbf{n}}_j)\sigma(\hat{\mathbf{n}}_i)\rangle=\sum_{\ell=0}^\infty\sum_{m=-\ell}^\ell C_{\ell m}Y^*_{\ell m}(\hat{\mathbf{n}}_i)Y_{\ell m}(\hat{\mathbf{n}}_j), \quad C_{\ell m}=\frac{1}{N^2}\left\lvert\sum_iA_i\sigma_iY_{\ell m}(\hat{\mathbf{n}}_i)\right\rvert^2,
\end{equation}
where $A_i$ denotes the area of the Voronoi cell associated with vertex $i$.
We choose the coordinate system in which the spherical harmonics $Y_{\ell m}(\hat{\mathbf{n}}_i)$ are defined such that the $z$-axis passes through two opposite vertices of the original regular icosahedron, thereby aligning it with a fivefold rotation axis.
Furthermore, the quantity $C_{\ell m}$ is a Monte Carlo estimator for the corresponding coefficient of the spin-spin correlator expansion and is averaged over configurations in the simulations (for a detailed derivation of the exact expansion, see Appendix A of~\cite{2024_brower_owen}).
The approximation is well-defined for angular momenta satisfying $\ell\ll \sqrt{N}$~\cite{2024_brower_owen}.

For a rotationally invariant system, the spin-spin correlator depends only on the angular separation between two points.
Consequently, all $2\ell+1$ coefficients $C_{\ell m}$ belonging to a fixed angular momentum $\ell$ must again be degenerate, which is the same criterion used to diagnose rotational symmetry in the free-particle spectrum in \cref{sec:latticeoptimization}.
For a general system with reduced symmetry, the coefficient $C_{\ell m}$ generally depends on $m$, reflecting the splitting of the continuum rotational multiplet.

This symmetry breaking can be quantified by comparing the minimum and maximum coefficients within a multiplet, defined as~\cite{2024_brower_owen}
\begin{equation}\label{eq:deltaCell}
    \delta C_{\ell}=1-\frac{\min_m(C_{\ell m})}{\max_m(C_{\ell m})},
\end{equation}
which vanishes when full degeneracy, and hence rotational invariance, is restored.

\begin{figure}[]
    \centering
    \begin{subfigure}[t]{0.49\textwidth}
        \centering
        \includegraphics[width=0.95\textwidth]{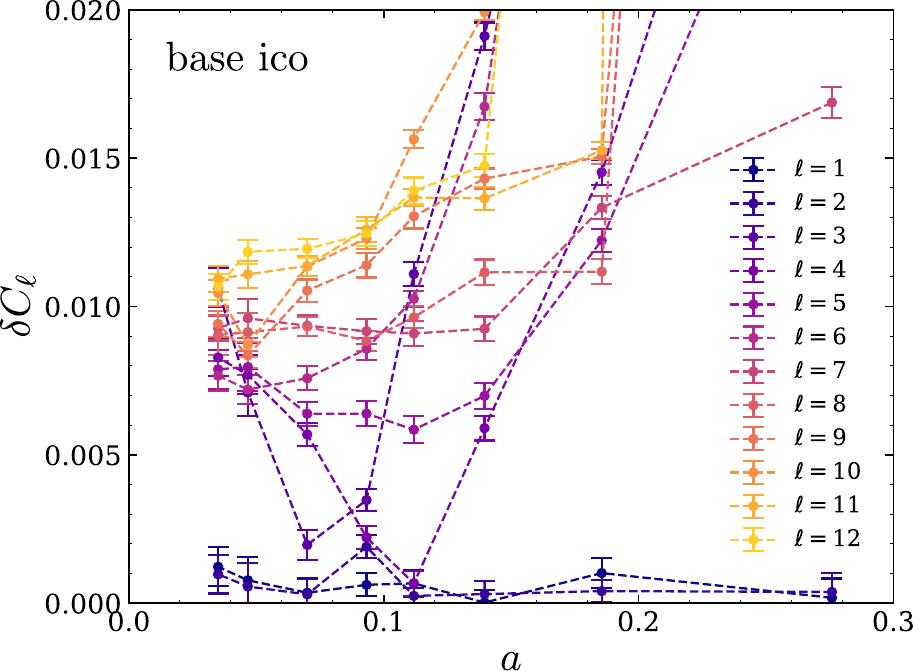}
    \end{subfigure}
    \begin{subfigure}[t]{0.49\textwidth}
        \centering
        \includegraphics[width=0.95\textwidth]{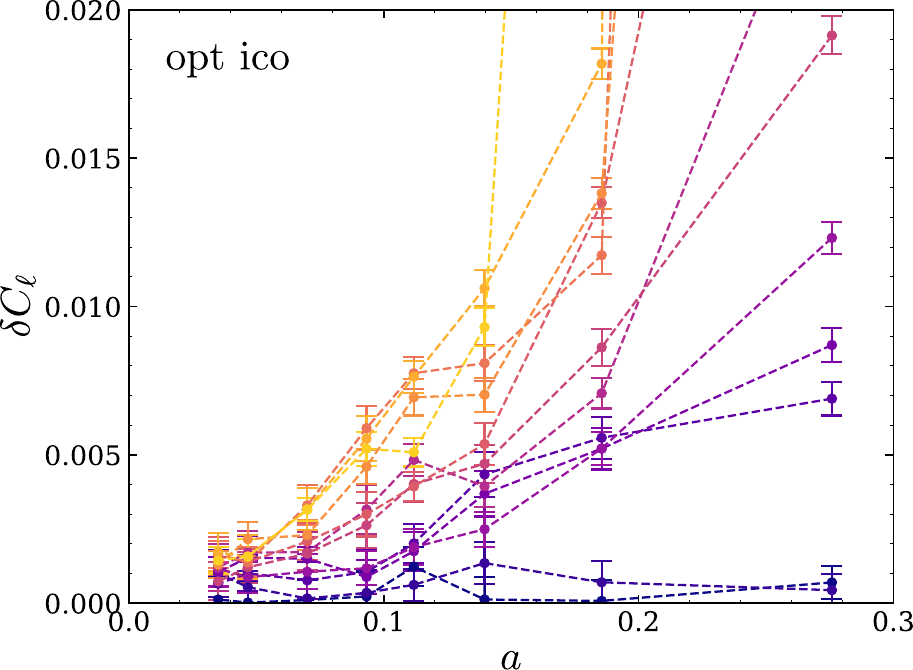}
    \end{subfigure}
    \caption{Measurement of the rotational symmetry breaking for the base (left) and the optimized (right) icosahedral lattices, shown as a function of the lattice spacing.}\label{fig:symmetrybreakingising}
\end{figure}
\cref{fig:symmetrybreakingising} depicts the symmetry breaking $\delta C_\ell$, with $1\le\ell\le12$, for both the base and the optimized icosahedral lattices with up to $N=10242$ vertices, obtained by running Monte Carlo simulations with $\mathcal{O}(10^8)$ sweeps (the statistical analysis was performed using the $\texttt{Carlo.jl}$ module~\cite{2025_Weber}).
For both lattice constructions, the coefficients with $\ell\le2$ remain degenerate within statistical uncertainty at all lattice spacings.
These are again the low-energy sectors which exhibited the expected spectral degeneracy in the previous section and could be explained by the decomposition of the continuum $O(3)$ representations under the residual icosahedral symmetry $I_h$, where $\ell=1,2$ representations remain irreducible.
We use these icosahedral-symmetry protected states to serve as a baseline for the expected behavior of a degenerate multiplet in the following analysis.

The higher angular momentum sectors which are not protected by the icosahedral symmetry provide a nontrivial test of rotational symmetry.
For $\ell>2$, the base icosahedral lattices exhibit a non-vanishing symmetry breaking in the continuum limit, indicating that the rotational symmetry of the interacting theory is not fully restored.
In contrast, for the optimized lattices, $\delta C_\ell$ decreases towards zero with increasing refinement for all measured $\ell$.
Thus, the lattice geometry obtained from spectral optimization combined with model-appropriate couplings restores rotational symmetry not only in the free-particle spectrum but also in the critical interacting Ising model.

Finally, we investigate the dependence of this result on the choice of couplings and temperatures (the corresponding results are presented in Appendix~\ref{sec:isingappendix}).
One could expect a spherically symmetric system to exhibit perfect spherical symmetry independent of the conditions it is put in.
However, we instead find that there is significant symmetry breaking for systems away from an RG fixed point, as well as for systems using Laplacian or constant couplings.
The restoration of rotational symmetry therefore depends on the combination of optimized lattice geometry and suitable critical couplings.

% !TeX root = ../SciPostPhys.txt

\section{Potts model on \texorpdfstring{$\mathbb{S}^2$}{S2}}
\label{sec:pottsmodelons2}
To test whether the restoration of rotational symmetry extends beyond the Ising universality class, we consider the critical $Q$-state Potts model, which is a natural choice here, since it is well known that the isoradial-graph construction again provides critical couplings~\cite{2018_Duminil_Copin}.
The Hamiltonian of the Potts model is given by
\begin{equation}
    H=-\sum_{\langle ij \rangle}J_{ij}\delta_{\sigma_i, \sigma_j}, \; \sigma_i\in[1, \dots, Q],
\end{equation} 
so that each Potts model configuration is weighted by the Boltzmann factor $\exp(-\beta H)$ at an inverse temperature $\beta=1/T$.
In two dimensions, the phase transition is continuous for $Q=2,3,4$, which allows us to study several distinct interacting critical theories on the same optimized lattice geometry.

The isoradial-graph critical couplings take the form~\cite{2018_Duminil_Copin}
\begin{equation}\label{eq:pottscouplings}
    K_{ij}=\beta J_{ij}=
    \begin{cases}
        \log\left(\frac{\sqrt{Q}\sin(r(\pi-\theta_{ij}))}{\sin(r\theta_{ij})}+1\right),
        &\text{if } Q=2, 3,\\
        \log\left(\frac{2\pi-\theta_{ij}}{\theta_{ij}}\right),
        &\text{if } Q=4,
    \end{cases}
\end{equation}
where $r=\arccos(\sqrt{Q}/2)/\pi$, and the rhombic angles $\theta_{ij}$ are defined as in the previous section.

For $Q=2$, the Potts couplings satisfy $K^{\text{Potts}}_{ij}=2K^{\text{Ising}}_{ij}$.
Upon proper rescaling, the $Q=2$ Potts model is then exactly equivalent to the Ising model up to an additive constant in the Hamiltonian.
We thus expect this case to reproduce the results of the previous section.

\subsection{Rotational symmetry of the Potts model on \texorpdfstring{$\mathbb{S}^2$}{S2}}
\label{subsec:rotsymmpottsons2}
We use the same base and spectrally optimized icosahedral lattices as in the Ising model, without further geometric optimization.
To quantify rotational symmetry breaking, we represent the Potts states by the $Q$-th roots of unity $q_i=\exp(2\pi i\sigma_i/Q)$~\cite{2020_Fujimoto}.
With that, the coefficients $C_{\ell m}$ become
\begin{equation}\label{eq:CellmPotts}
        C_{\ell m} = \frac{1}{N^2}\left\lvert \sum_i A_i q_i Y_{\ell
        m}(\hat{\mathbf{n}}_i)\right\rvert^2,
\end{equation}
which we use to evaluate the symmetry-breaking measure $\delta C_\ell$ defined in \cref{eq:deltaCell}.
This way, the representation of the Potts states as roots of unity provides a complex parameter whose spherical-harmonic modes can be analyzed using the same procedure as for the Ising spin field.

\begin{figure}[h!]
    \centering
    \begin{subfigure}[t]{0.49\textwidth}
    \centering
        \includegraphics[width=0.95\textwidth]{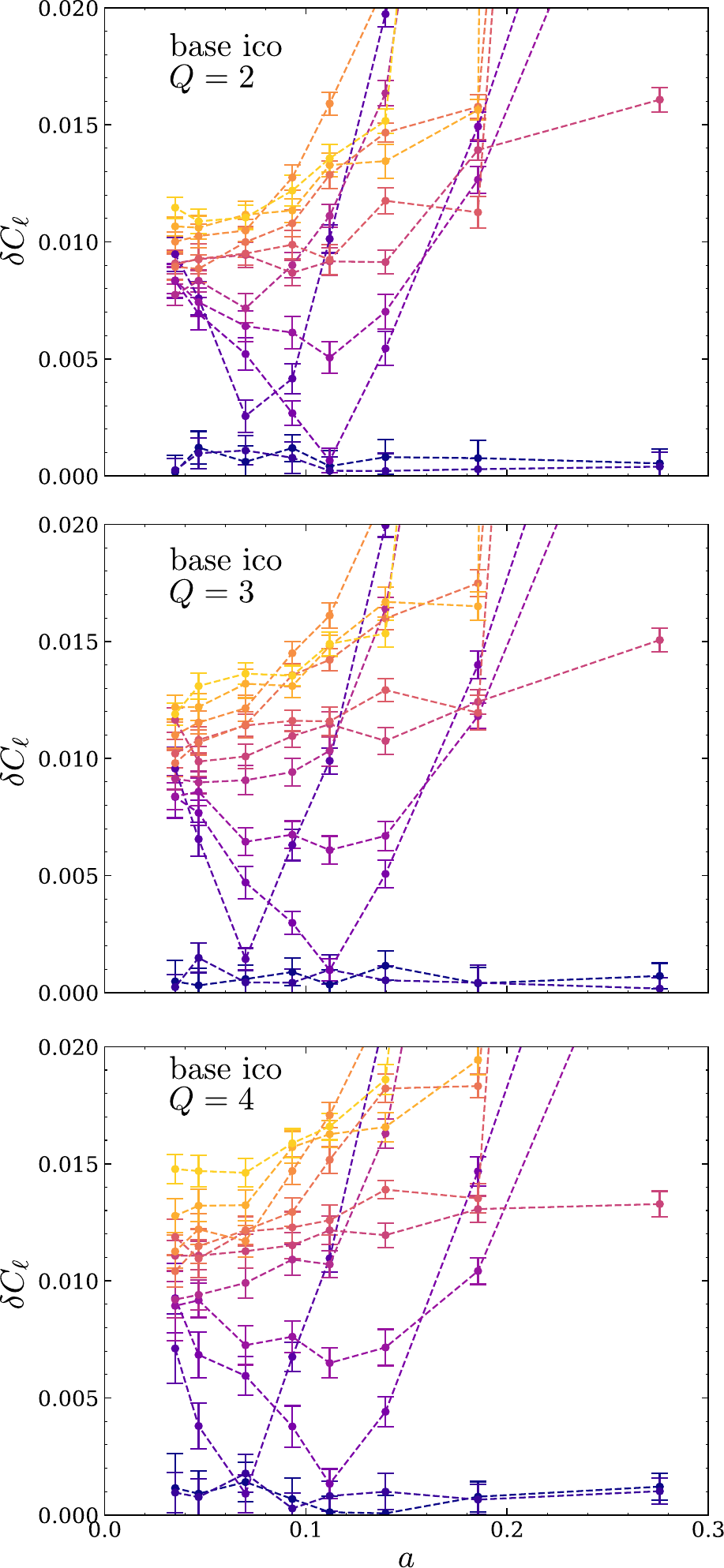}
    \end{subfigure}
    \begin{subfigure}[t]{0.49\textwidth}
    \centering
        \includegraphics[width=0.95\textwidth]{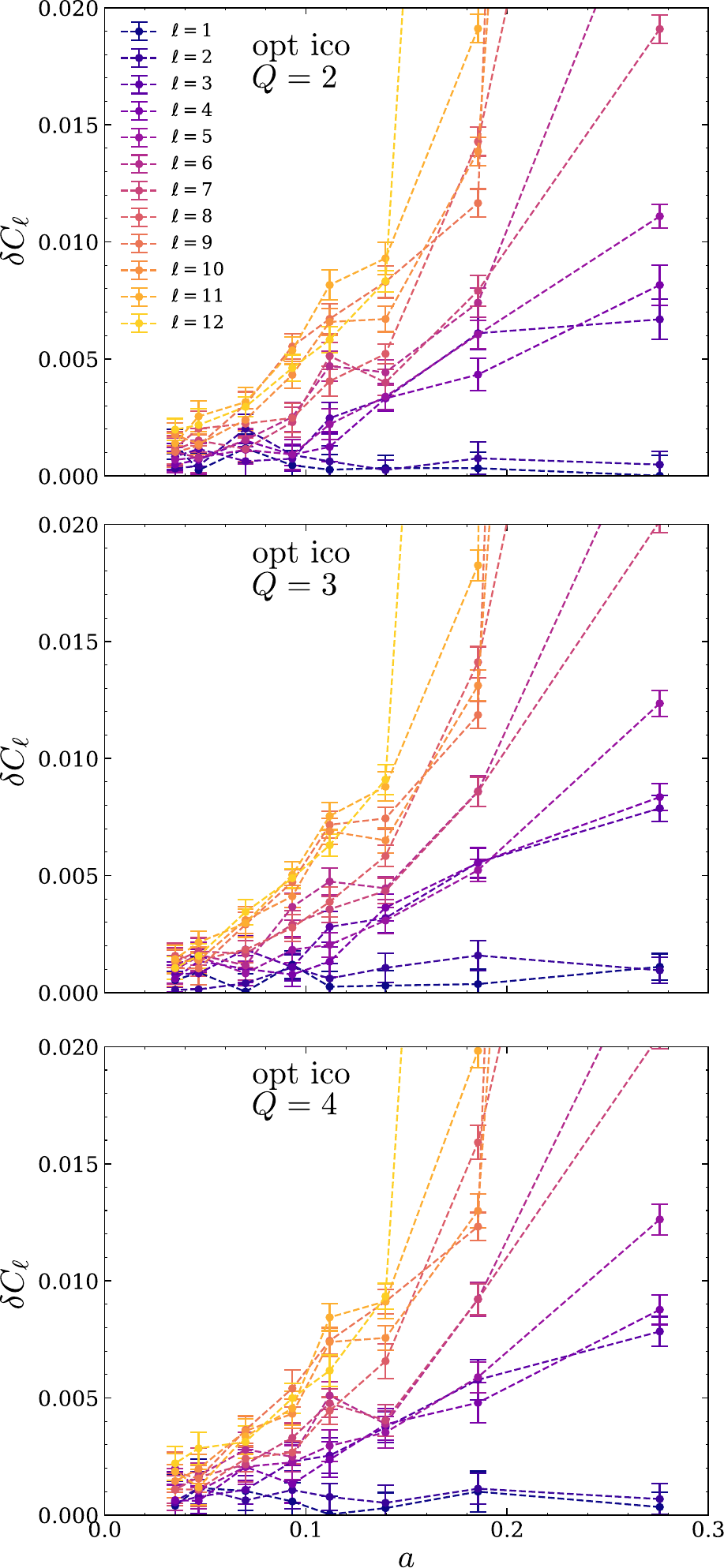}
    \end{subfigure}
    \caption{Measurement of the rotational symmetry breaking for the base (left) and the optimized (right) icosahedral lattices.
    The rows correspond to $Q=2,3,4$.}\label{fig:symmetrybreakingpotts}
\end{figure}

Monte Carlo simulations for $Q=2,3,4$ on the base and optimized icosahedral lattices give the results shown in \cref{fig:symmetrybreakingpotts}.
For $Q=2$, the expectation due to the exact equivalence to the Ising model is fulfilled; the base lattices retain a nonzero rotational symmetry breaking in the continuum limit, whereas $\delta C_\ell$ decreases towards zero for the optimized lattices.
The agreement with the Ising results provides a consistency check for the Potts construction.

For $Q=3$ and $Q=4$, we observe the same qualitative behavior.
The symmetry-breaking measure decreases towards the baseline given by $\ell=1,2$ with increasing refinement for the optimized lattices, while remaining substantially larger for the base lattices.
Thus, the restoration of rotational symmetry induced by the optimized geometry is not specific to the Ising model, but persists across other interacting critical theories, given a suitable choice of critical couplings.

% !TeX root = ../SciPostPhys.tex 

\section{Transverse field Ising model on \texorpdfstring{$\mathbb{S}^2$}{S2}}
\label{sec:tfimons2}
Having demonstrated the restoration of rotational symmetry for classical critical models, we now proceed by extending this analysis to the transverse-field Ising model (TFIM), whose zero-temperature quantum phase transition allows us to test whether the optimized geometry also reproduces the universal properties of an interacting quantum critical point.
The Hamiltonian of the TFIM is given by
\begin{equation}
    \hat{H} = -\sum_{\langle ij \rangle} J_{ij} \hat{S}^z_i \hat{S}^z_j - h\sum_i \hat{S}^x_i,
\end{equation}
with ferromagnetic couplings $J_{ij}>0$ and a transverse field in $x$-direction with field strength $h\ge0$. The quantum partition function at inverse temperature $\beta=1/T$ is given by $Z=\mathrm{Tr} \:e^{-\beta \hat H} $.
At $h=0$, the model has a ferromagnetically ordered ground state, while for $h\gg J_{ij}$ the system approaches a field-polarized paramagnet.
Between these two limits, it undergoes a continuous quantum phase transition at a critical field $h_c$, associated with a spontaneous breaking of the $\mathbb{Z}_2$ symmetry generated by the parity operator $\hat{P}=\prod_i\hat{\sigma}^x_i$, where $\hat{S}^x = \frac{\hbar}{2}\hat{\sigma}^x$.
To characterize the phase transition, we use the spontaneous magnetization in the $z$-direction $\hat{m}=\frac{1}{N}\sum_i\hat{S}^z_i$ as the order parameter operator.

Before investigating rotational symmetry at the quantum critical point, we first identify the critical transverse field of the TFIM on the optimized lattices.
Here, the couplings $J_{ij}$
are chosen proportional to the Ising couplings defined in \cref{eq:isingcouplings}. More specifically, we set
\begin{equation}\label{eq:isingcouplingsQ}
J_{ij}=\frac12\log\left(\frac{1+\cos(\theta_{ij}/2)}{\sin(\theta_{ij}/2)}\right).
\end{equation}
This choice is motivated by the fact that the limiting case $h\to0$ effectively reduces the TFIM to the classical Ising model, where these couplings place the system at criticality upon tuning the temperature to $T=1$.
Away from this limit, we anticipate an optimal choice of couplings to be nontrivially dependent on the transverse field.
Such a dependence is, however, not known at present and remains an interesting direction for future work.

\subsection{Locating the critical point via finite-size scaling}
\label{sec:tfimcriticality}
We determine $h_c$ using crossings of the Binder cumulant~\cite{1981_Binder}
\begin{equation}\label{eq:bindercumulant}
    U_4 = \frac{3}{2} \left(1-\frac13\frac{\langle \hat{m}^4\rangle}{\langle \hat{m}^2\rangle^2}\right),
\end{equation}
which, written in this form, has the thermodynamic limits $U_4\to1$ in the ordered and $U_4\to0$ in the disordered phase, respectively.
Near the phase transition, the Binder cumulant obeys a finite-size scaling form of the type $U_4(h, L)=f_U\left((h-h_c)L^{1/\nu}\right)$, with the linear system size $L$ and the critical exponent $\nu$.
This implies that at $h=h_c$ it becomes independent of the system size up to corrections to scaling, enabling the crossing analysis.
Since the surface area is fixed, the linear system size of our spherical lattices scales as $L\sim\sqrt{N}$.
To perform the calculations, we use stochastic series expansion~\cite{2003_Sandvik} quantum Monte Carlo simulations at inverse temperatures $\beta=2\sqrt{N}/J$, with $J=\langle J_{ij}\rangle\approx0.276$, to ensure convergence to the ground state of the system 
(we verified that further increasing $\beta$ does not change the measured observables within statistical uncertainty).

\begin{figure}[h!]
    \centering
    \includegraphics[width=0.85\textwidth]{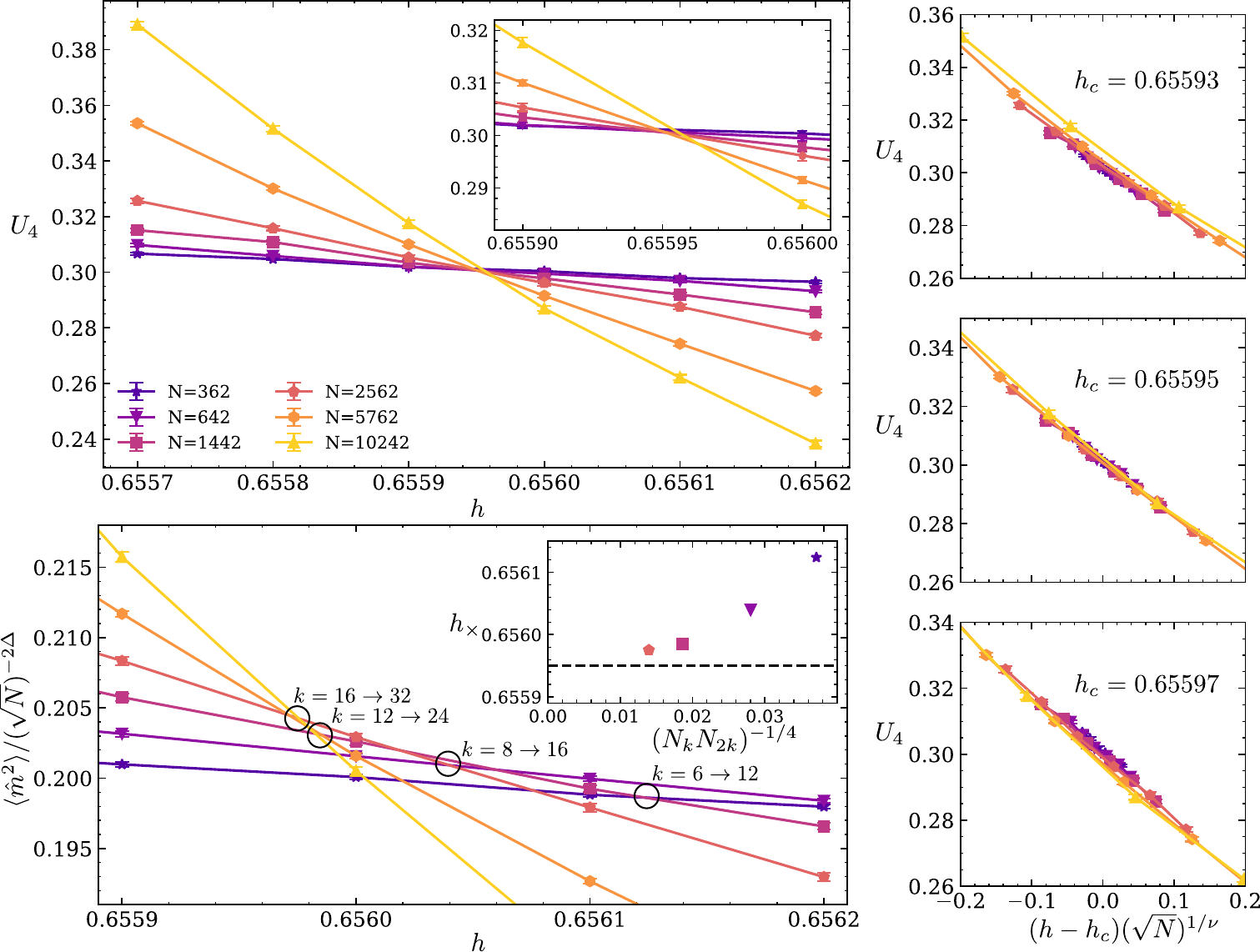}
    \caption{Upper left: Binder cumulants of different system sizes shown as a function of the transverse field.
    Right: Collapse of the Binder cumulants for different values of $h_c$ with respect to $(h-h_c)(\sqrt{N})^{1/\nu}$, where $\nu=0.63002$~\cite{2010_Hasenbusch}.
    Lower left: Crossings of the order parameter rescaled with the appropriate finite-size scaling form $\langle \hat{m}^2\rangle/(\sqrt{N})^{-2\Delta}$, where $\Delta=0.518151$~\cite{2015_simmonsduffin}.
    The inset shows the crossing values $h_\times$ with respect to the inverse geometric mean of the linear system sizes involved, i.e. $(N_6N_{12})^{-1/4}$ for the crossing $k=6\to 12$.
    The black dashed line indicates the critical transverse field $h_c=0.65595$.}\label{fig:bindercumulants}
\end{figure}

In the upper left corner of \cref{fig:bindercumulants}, we show the Binder cumulants with respect to the transverse field for systems with up to $N=10242$ vertices.
All curves cross within windows of $h\in[0.65593, 0.65597]$ and $U_4\in[0.298, 0.302]$, from which we determine the location of the critical point:
\begin{equation*}
    h_c=0.65595\pm0.00002 \quad \text{and}\quad U^c_4=0.3\pm0.002,
\end{equation*}
where we conservatively estimate the uncertainty as half the width of the crossing window.
The critical value of the Binder cumulant is consistent with the estimate $0.28\le U_4^c\le0.4$ presented in a recent paper using the fuzzy sphere regularization method~\cite{2023_Zhu}.

We also show the collapse of the Binder cumulant curves when plotted with respect to the finite-size scaling form $(h-h_c)(\sqrt{N})^{1/\nu}$, where $\nu\approx0.63002$ is a well-known exponent of the $3\mathrm{D}$ Ising universality class~\cite{2010_Hasenbusch}, on the right side of \cref{fig:bindercumulants}.
For $h_c=0.65595$, we observe a clean collapse, providing further confirmation of the estimated critical field.

As a final consistency check, we investigate the crossings in the curves of the squared order parameter $\langle \hat{m}^2\rangle$, rescaled with the appropriate finite-size relation $1/(\sqrt{N})^{-2\Delta}$, with the scaling dimension of the Ising order parameter $\Delta\approx0.518151$~\cite{2015_simmonsduffin}.
Looking at the plot shown in the bottom left corner of \cref{fig:bindercumulants}, we find that the curves exhibit a converging behavior from the right to the critical transverse field strength determined earlier using the Binder cumulants.

\subsection{Conformal symmetry at criticality}
\label{subsec:conformalsymmetry}
Having located the system's critical point, we proceed to test whether the spectrum of the TFIM on our finite spherical lattices is consistent with the operator content of the $3\mathrm{D}$ Ising conformal field theory.
Before presenting the results, we briefly establish the theoretical background that enables them.

First, the quantum critical point of the TFIM in two spatial dimensions belongs to the $3\mathrm{D}$ Ising universality class.
It is therefore expected that the long-wavelength theory at the critical point is described by the $3\mathrm{D}$ Ising CFT.
To relate the spectrum of the TFIM critical theory to CFT data, it is convenient to consider the quantization of the CFT on the cylinder $\mathbb{S}^2\times\mathbb{R}$, which coincides with the geometry of the path-integral representation of the TFIM on $\mathbb{S}^2$.

In general, a quantization of a CFT defined on flat Euclidean space $\mathbb{R}^3$ can be mapped to such a cylinder via a Weyl transformation, known as radial quantization.
Under this mapping, the logarithm of the radial coordinate in $\mathbb{R}^3$ becomes the imaginary-time coordinate and the unit sphere surrounding the origin becomes a spatial slice of the cylinder~\cite{1984_Cardy, 1985_Cardy}.
A central consequence of radial quantization is the state-operator correspondence: if a critical theory has conformal symmetry, local operators of the CFT correspond to states in its quantization on $\mathbb{S}^2$.
In particular, the energy gaps of the eigenstates of the corresponding quantum Hamiltonian can be related to the scaling dimensions of the associated CFT operators~\cite{1984_Cardy, 1985_Cardy}:
\begin{equation}
    \Delta E_n=E_n-E_0=\frac{c}{R}\Delta_n,
\end{equation}
where $R\propto\sqrt{N}$ is the radius of the sphere, $c$ is a nonuniversal characteristic velocity, and $\Delta_n$ is the scaling dimension of the CFT operator associated with the $n$-th eigenstate.
Thus, the slopes of the finite-size scaling of the energy gaps of our TFIM systems determine the products $c\Delta_n$, while ratios of energy gaps directly yield ratios of scaling dimensions independent of the nonuniversal velocity $c$.

To apply this to specific operators, it is useful to know that the operators of a CFT are organized into conformal multiplets consisting of a primary operator and its descendants~\cite{2016_simmonsduffin}.
Primary operators transform covariantly under the conformal group, while descendants are generated by acting with derivatives on the primary and therefore have scaling dimensions shifted by integer amounts.
Here, we focus on the scalar primary operators $\sigma$ and $\epsilon$ with scaling dimensions $\Delta_\sigma$ and $\Delta_\epsilon$.
Their $\mathbb{Z}_2$ parity distinguishes the two operators: $\sigma$ is odd, whereas $\epsilon$ is even.
The corresponding lowest-energy states therefore lie in the odd and even parity sectors of the Hamiltonian.
Importantly, $\sigma$ and $\epsilon$ have quantum number $\ell=0$, while their descendants are not scalar; instead they correspond to spin-$\ell$ representations with $\ell \geq 0$.

We can thus test the consistency of the critical lattice theory with the $3\mathrm{D}$ Ising CFT by verifying the expected state-operator correspondence and comparing ratios of energy gaps with the ratio of the following known scaling dimensions obtained from conformal bootstrap calculations~\cite{2015_simmonsduffin}:
\begin{equation*}
    \Delta_\sigma\approx0.518151, \qquad \Delta_\epsilon\approx1.41264.
\end{equation*}

\subsubsection{Accessing the excitation gaps}
To access the excitation gaps of the TFIM on our spherical lattices, we measure imaginary-time correlation functions of suitable operators $\hat{A}$.
The spectral decomposition of such a correlator receives contributions from all excited states with nonzero overlap with $\hat{A}$.
At sufficiently low temperature, $\beta\Delta E_1\gg1$, we can assume that the thermal occupation of excited states becomes negligible, so the system is effectively in its ground state.
Then, the correlator takes the form~\cite{2012_Hen}
\begin{equation}\label{eq:correlatorScalingDims1}
    C_{\hat{A}}(\tau)=\langle\hat{A}(\tau)\hat{A}(0)\rangle-\langle\hat{A}\rangle^2\approx\sum_{n=1}\lvert\langle0\lvert\hat{A}\rvert n\rangle\rvert^2\left(e^{-\Delta E_n\tau}+e^{-\Delta E_n(\beta-\tau)}\right),
\end{equation}
where the second exponential term accounts for the periodic boundary conditions in imaginary time and ensures symmetry around $\tau=\beta/2$.
If we additionally assume that $\tau$ is sufficiently large for contributions from higher excited states to decay away faster than the lowest gap, we can further approximate the correlator as~\cite{2012_Hen}
\begin{equation}\label{eq:correlatorScalingDims2}
    C_{\hat{A}}(\tau)\approx\lvert\langle0\lvert\hat{A}\rvert 1\rangle\rvert^2e^{-\Delta E_1\tau}.
\end{equation}
A necessary condition for extracting the lowest gap in a given symmetry sector is therefore a nonzero overlap $\lvert\langle0\lvert\hat{A}\rvert 1\rangle\rvert$.
To extract the gap associated with the odd primary $\sigma$, we use the odd magnetization operator $\hat{m}=\frac{1}{N}\sum_i\hat{S}_i^z$,
while for the even primary $\epsilon$ we use the energy operator $\hat{E}=\sum_{\langle ij\rangle}J_{ij}\hat{S}_i^z\hat{S}_j^z$.
We extract the corresponding gaps from the logarithmic decay of the correlation functions in a time window where the contribution of the lowest state dominates, while remaining sufficiently far from $\tau=0$ and $\tau=\beta/2$ to suppress short-time effects and retain an adequate signal-to-noise ratio.

\begin{figure}[]
    \centering
    \includegraphics[width=0.7\textwidth]{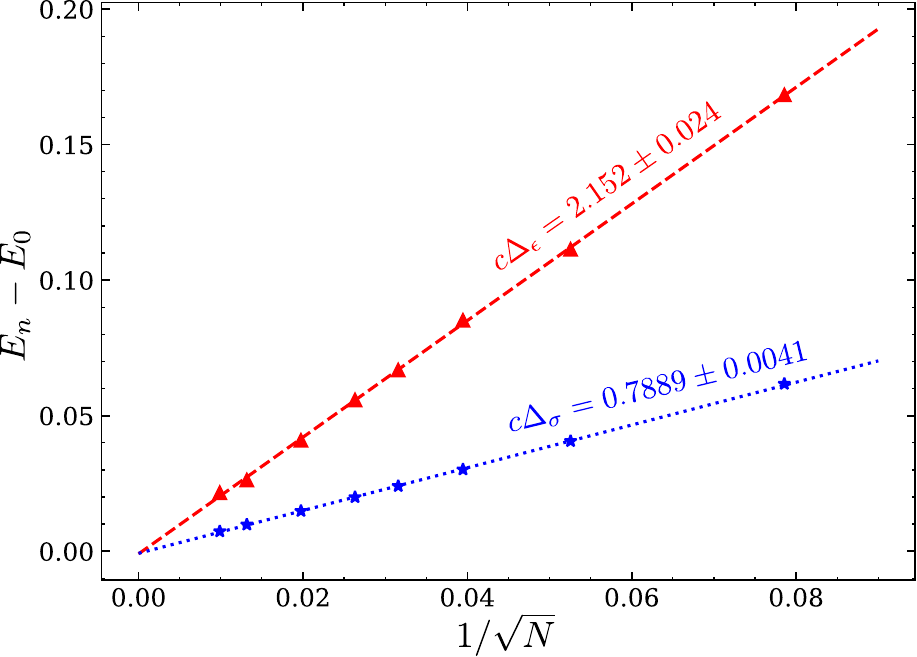}
    \caption{Linear fit of the energy gaps in the odd (blue, dotted) and even (red, dashed) parity sectors with respect to the inverse linear sizes of the systems.}\label{fig:primaryscalingdims}
\end{figure}
The lowest energy gaps in the odd and even parity sectors are shown as a function of $1/\sqrt{N}$ in \cref{fig:primaryscalingdims}.
Both exhibit good agreement with the expected linear finite-size scaling form, with $\chi^2/n_\text{dof}=1.3$ for $c\Delta_\sigma$ and $\chi^2/n_\text{dof}=0.6$ for $c\Delta_\epsilon$.
From the slopes, we obtain
\begin{equation*}
    c\Delta_\sigma=0.7889\pm0.0041,
    \qquad
    c\Delta_\epsilon=2.152\pm0.024.
\end{equation*}
The quoted uncertainties are obtained from the least-squares finite-size scaling fits and do not include systematic uncertainties associated with the choice of the decay window in the preceding correlator fits.
Taking the ratio of the extracted products eliminates the nonuniversal velocity $c$ and gives
\begin{equation*}
    \frac{\Delta_\sigma}{\Delta_\epsilon}=0.3666\pm0.0046.
\end{equation*}
Our result is consistent within one standard deviation with the conformal bootstrap prediction $\Delta_\sigma/\Delta_\epsilon=0.3668$~\cite{2015_simmonsduffin} and with the recent fuzzy-sphere result $\Delta_\sigma/\Delta_\epsilon=0.3709$~\cite{2023_Zhu}.
The agreement provides evidence that the low-lying excitation spectrum of the critical TFIM on the optimized spherical lattices is consistent with the spectrum of the $3\mathrm{D}$ Ising CFT.
%We next test whether this agreement extends beyond the scalar sector to descendants with nonzero angular momentum.

\subsubsection{Angular-momentum resolved excitation gaps at criticality}
The scalar primary operators with zero angular momentum, whose state-operator correspondence we were able to show above, cannot resolve the splitting of rotational multiplets induced by the reduced symmetry of the lattice, since this sector contains only a single angular-momentum component.
To probe rotational symmetry beyond the scalar sector, we therefore study descendants of the primary operator $\sigma$ corresponding to $\ell>0$ states.

In a conformal multiplet, descendants obtained by acting with $\ell$ derivatives on a scalar primary have scaling dimensions~\cite{2016_simmonsduffin}
\begin{equation}
    \Delta_{\sigma, \ell}=\Delta_\sigma+\ell.
\end{equation}
We access these states by projecting the spin operator onto real spherical harmonics, yielding $\hat{m}'_{\ell, m}=\frac{1}{N}\sum_iA_iY_{\ell, m}(\hat{\mathbf{n}}_i)\hat{S}^z_i$.
This operator has nonzero overlap with the lowest-energy state in the odd sector carrying angular momentum quantum numbers $(\ell, m)$.
It therefore allows us to extract the scaling dimensions of descendants of $\sigma$, again using a correlator of the form given in \cref{eq:correlatorScalingDims1}.

\begin{figure}
    \centering
    \begin{subfigure}[t]{0.51\textwidth}
        \centering
        \includegraphics[width=0.95\textwidth]{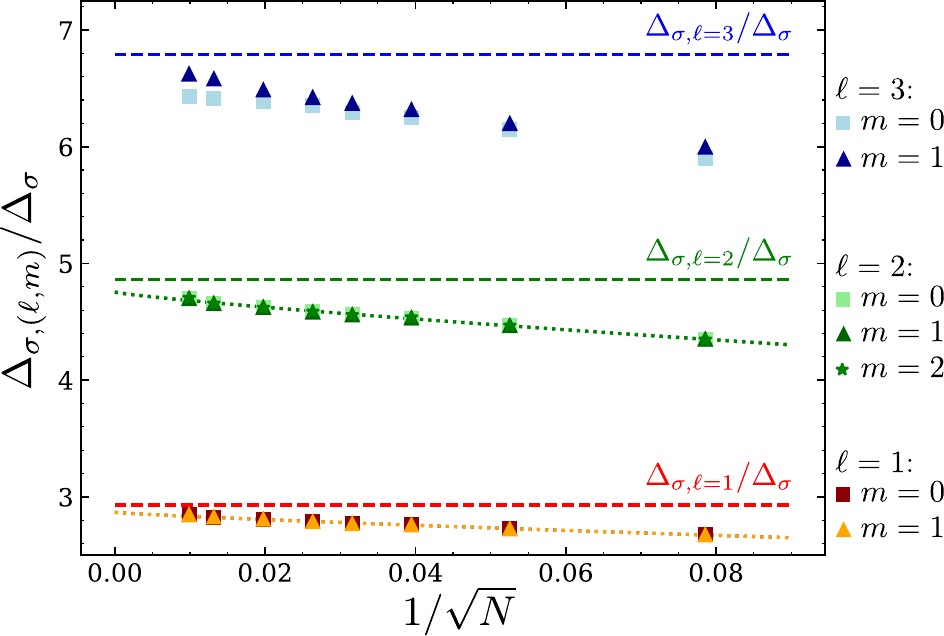}
    \end{subfigure}
    \begin{subfigure}[t]{0.465\textwidth}
        \centering
        \includegraphics[width=0.95\textwidth]{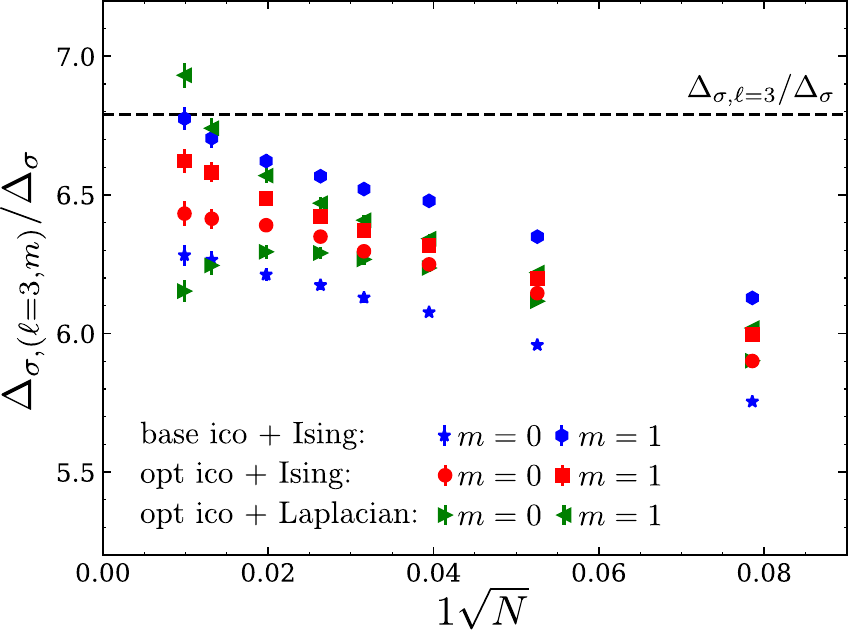}
    \end{subfigure}
    \caption{Left: ratios of scaling dimensions $\Delta_{\sigma,(\ell, m)}/\Delta_{\sigma}$ for $\ell=1, 2, 3$ using the optimized lattices and Ising couplings.
    Dashed lines indicate the expected CFT operator scaling dimension ratio values.
    Dotted lines indicate the finite-size scaling form fit with $(\Delta_{\sigma, (\ell, m)}/\Delta_{\sigma})(L)=(\Delta_{\sigma, (\ell, m)}/\Delta_{\sigma})_\infty+AL^{-\omega}$, where $\omega=0.832$~\cite{2010_Hasenbusch}.
    Right: ratios of scaling dimensions $\Delta_{\sigma,(\ell=3, m)}/\Delta_{\sigma}$ for the combinations i) optimized lattices and Ising couplings (red), ii) base lattices and Ising couplings (blue), iii) optimized lattices and Laplacian couplings (green). }\label{fig:descendantscalingdims}
\end{figure}

In the left panel of \cref{fig:descendantscalingdims}, we show the ratios $\Delta_{\sigma, (\ell, m)}/\Delta_{\sigma}$ for $\ell=1, 2, 3$ as a function of $1/\sqrt{N}$, extracted by using operators projected onto spherical harmonics with different quantum numbers $m$.
For $\ell=1$ and $\ell=2$, the different $m$ components give statistically consistent values, indicating that the corresponding descendant multiplets remain degenerate, as was to be expected from the underlying icosahedral symmetry.
We perform finite-size scaling on these ratios using the functional form $(\Delta_{\sigma, (\ell, m)}/\Delta_{\sigma})(L)=(\Delta_{\sigma, (\ell, m)}/\Delta_{\sigma})_\infty+AL^{-\omega}$, where $\omega\approx0.832$ is the leading correction-to-scaling exponent for the $3\mathrm{D}$ Ising universality class~\cite{2010_Hasenbusch}.
The resulting continuum extrapolations are summarized in \cref{tab:scalingdims}.
The extrapolated ratios are again statistically consistent across all $m$ components, while differing from the expected CFT values by at most $2.2\%$, with a small systematic shift towards smaller values.
This systematic shift may be related to finite-size effects and resolving them would require a deeper exploration of the finite-size regime.
Nevertheless, together with the agreement of the primary scaling dimensions, this provides evidence that the optimized lattices reproduce the expected descendant spectrum in the symmetry-protected $\ell=1$ and $\ell=2$ sectors.
\begin{table}[h!]
	\centering
	\begin{tabular}{c|c|c|c|c}\toprule
        \multicolumn{2}{c|}{Quantum numbers} & FSS value & Bootstrap value & Rel. error \\
		\midrule
		\multirow{2}{*}{$\ell=1$}  & $m=0$ & $2.8681\pm0.0086$ & \multirow{2}{*}{$2.9299$} & $2.1\%$ \\
        & $m=1$ & $2.8681\pm0.0086$ & & $2.1\%$ \\ \hline
		\multirow{3}{*}{$\ell=2$}  & $m=0$ & $4.7560\pm0.0133$ & \multirow{3}{*}{$4.8599$} & $2.1\%$ \\
        & $m=1$ & $4.7522\pm0.0132$ & & $2.2\%$ \\
        & $m=2$ & $4.7567\pm0.0135$ & & $2.1\%$ \\ \bottomrule
	\end{tabular}
	\caption{Scaling dimension ratios $(\Delta_{\sigma, (\ell, m)}/\Delta_{\sigma})_\infty$ extracted via finite-size scaling for $\ell=1, m=0,1$ and $\ell=2, m=0, 1, 2$.
    We also show the expected value $(\Delta_\sigma+\ell)/\Delta_\sigma$, where $\Delta_\sigma=0.518151$~\cite{2015_simmonsduffin} is the bootstrap value, and the relative error.}\label{tab:scalingdims}
\end{table}

For $\ell=3$, we explicitly only show the sectors with $m=0$ and $m=1$.
Due to the alignment of the fivefold rotation axis of the lattice with the $z$-axis, the continuum spherical-harmonic components with $m=\pm1$ at $\ell=3$ project exclusively onto the $G_u$ irrep of $I_h$, while the $m=0$ component projects exclusively onto the $T_{2u}$ irrep.
The remaining components with $m=\pm2,\pm3$ contain contributions from both irreps.
Consequently, the $m=0$ and $m=1$ components provide a direct comparison between two states that are degenerate in the continuum but belong to distinct irreducible representations of the lattice symmetry group.

In contrast to the $\mathrm{SO}(3)$ restoration we observed for the classical models and for $\ell=1$ and $\ell=2$ in the quantum case, for $\ell=3$ there is a noticeable gap between measurements with $m=0$ and $m=1$, indicating that there is residual symmetry breaking that does not vanish in the continuum limit.
Instead, the gap increases as the lattice spacing decreases, suggesting a perturbation that has large effects for small lattice spacings.
The combination of optimized lattices and classical Ising couplings is thus not sufficient to restore spherical symmetry in the quantum case and a more nuanced approach is needed.

In addition to the optimized lattices with the critical Ising couplings, we also consider two reference constructions at their respective critical points: the optimized lattices with Laplacian couplings (at $h_c=0.6929\pm0.00005$) and the base lattices with Ising couplings (at $h_c=0.65515\pm0.00005$).
This comparison reinforces the observation that the optimized geometry and the appropriate couplings are two distinct requirements.

The results are shown in the right panel of \cref{fig:descendantscalingdims}.
We observe that the gap between the measurements using $m=0$ and $m=1$ exists for all three combinations.
Therefore, none of the three achieves a restoration of full rotational symmetry.
Nevertheless, the combination of optimized lattices and critical Ising couplings consistently exhibits the smallest splitting between the two $\ell=3$ components and is therefore the closest approximation to $\mathrm{SO}(3)$ for the TFIM.

\subsection{Rotational symmetry of the TFIM on \texorpdfstring{$\mathbb{S}^2$}{S2}}
\begin{figure}[t]
    \centering
    \begin{subfigure}[t]{0.49\textwidth}
        \centering
        \includegraphics[width=0.95\textwidth]{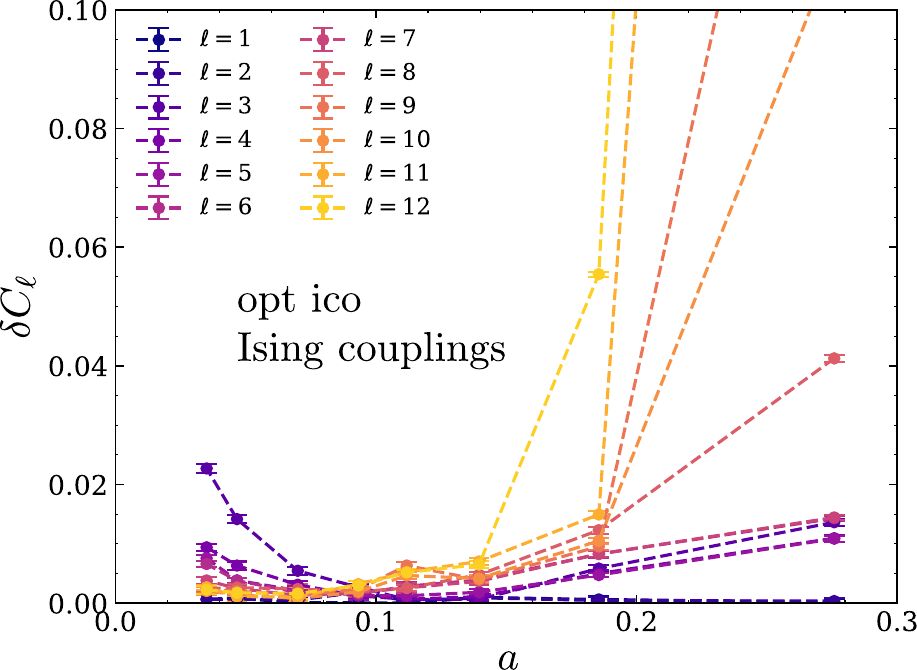}
    \end{subfigure}
    \begin{subfigure}[t]{0.49\textwidth}
        \centering
        \includegraphics[width=0.95\textwidth]{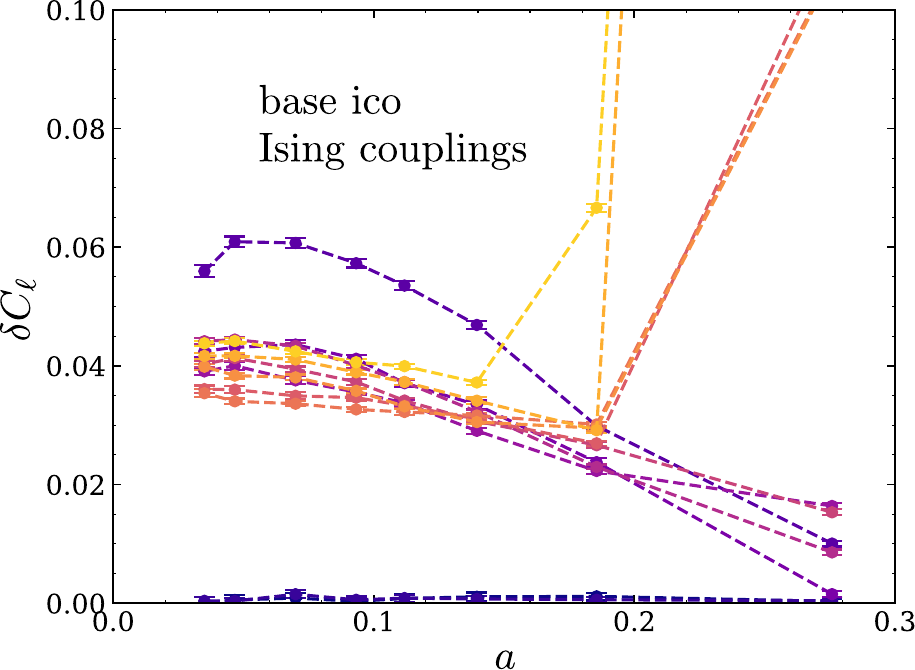}
    \end{subfigure}
    \begin{subfigure}[t]{0.49\textwidth}
        \centering
        \includegraphics[width=0.95\textwidth]{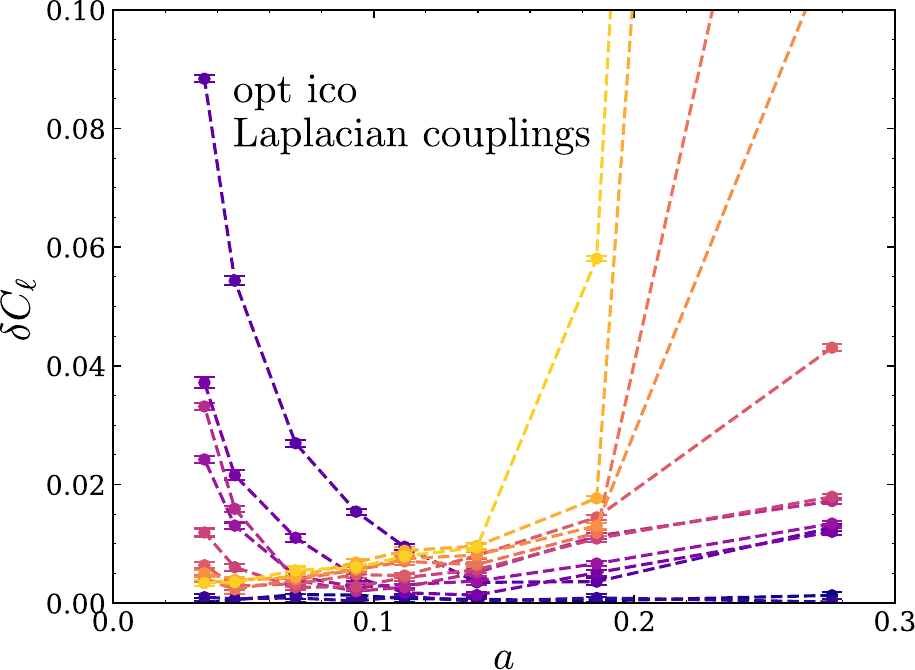}
    \end{subfigure}
    \caption{Top left: rotational symmetry breaking for the TFIM on the optimized lattices using Ising couplings.
    Top right: rotational symmetry breaking for the TFIM on the base lattices using Ising couplings.
    Bottom: rotational symmetry breaking for the TFIM on the optimized lattices using Laplacian couplings.}\label{fig:symmetrybreakingtfim}
\end{figure}

Finally, we repeat the rotational symmetry breaking measurement that we applied in the classical models to support the findings from the previous subsection.
Shown in \cref{fig:symmetrybreakingtfim}, we find a functionally equivalent picture to the $\Delta_{\sigma, (\ell, m)}/\Delta_{\sigma}$ ratio plots; there is a vanishing symmetry breaking for $\ell=1, 2$, while for $\ell\ge3$ the symmetry breaking increases for fine lattice refinements.
The combination of optimized lattices and Ising couplings provides again the closest approximation to $\mathrm{SO}(3)$ symmetry.
Interestingly, while $\delta C_\ell$ initially decreases for measurements using the optimized icosahedral lattices, it monotonically increases for measurements using base icosahedral lattices and Ising couplings for $2<\ell<8$ up to a system size of $N=2562$.

\section{Conclusion}
We have developed a gradient descent optimization algorithm for spherical lattices based on the degeneracy structure of the discretized Laplacian.
Although the optimization is defined in terms of the free-particle spectrum, it naturally drives refined icosahedral lattices towards geometries that exhibit both substantially improved rotational symmetry and geometric uniformity.

Using suitable couplings from the isoradial construction, enabled by the asymptotic isoradiality of the optimized lattices, we find that the optimized lattices recover $\mathrm{SO}(3)$ symmetry in the continuum limit for the Ising and Potts models.
For the Ising model, we furthermore find agreement with the geometric uniformity and critical coupling construction obtained by Brower and Owen~\cite{2024_brower_owen}.

Furthermore, we investigated the quantum phase transition of the TFIM, determining the critical transverse field and the critical Binder cumulant with high precision, and extracting excitation gaps in the even and odd parity sectors.
The resulting ratio of the primary scaling dimensions $\Delta_\sigma/\Delta_\epsilon=0.3666\pm0.0046$ is consistent with the $3\mathrm{D}$ Ising CFT and provides evidence for the state-operator correspondence on the lattice.
However, extending this analysis to descendant operators of $\sigma$ with angular momentum quantum number $\ell>0$, we found that only the symmetry-protected states with $\ell=1,2$ approximately recover the expected ratio of scaling dimensions and exhibit no statistically significant splitting between the different $m$ components.
Instead, the $\ell=3$ multiplet exhibits a splitting between states belonging to different irreducible representations of the underlying icosahedral symmetry.
Thus, while the optimized lattices substantially reduce rotational symmetry breaking, they do not completely eliminate lattice artifacts in higher-angular-momentum sectors.
We also note that Ising couplings provide a better approximation to $\mathrm{SO}(3)$ symmetry than Laplacian couplings, which indicates that their form more closely approximates the correct coupling construction needed.
As already mentioned, we assume that the correct couplings depend nontrivially on the transverse field, which is not yet known and is therefore a natural direction for future work.
Instead of adjusting the couplings, we tried using a non-constant transverse field, adjusting $h$ with the Voronoi cell area of each vertex, which however resulted in increased spherical symmetry breaking.

With regard to future research directions, we note the interesting connection between lattice geometry and the couplings of the interacting theories studied.
As found in~\cite{2023_Brower_affine}, the critical couplings for the Ising model on an affine triangular lattice are determined directly from the corresponding Laplacian couplings using $\sinh({2\beta J_{ij}}) = l^*_{ij}/l_{ij}$.
On the spherical geometry, each bond is shared by two distinct triangles.
We assigned to each bond a coupling obtained by averaging the values prescribed to the two adjacent triangles and found coupling strengths equal to the ones derived in~\cite{2024_brower_owen}, up to machine precision.
This approach may also provide a useful strategy for studying other interacting theories on the sphere.
Namely, one may seek to find a mapping between local lattice geometry and the critical couplings of the interacting theory and then proceed with a similar averaging prescription.
Such a mapping need not be known analytically.
Instead, it could be established first numerically on a regular/affine lattice and subsequently used to define the couplings on the spherical lattice. It would also be valuable to understand the origin of our finding of the emerging geometric uniformity of the spectrally optimized spherical lattices. 

Lastly, it would be interesting to investigate whether analogous optimization strategies to ours can be applied to other interacting lattice models and to spherical geometries in higher dimensions.

The optimization code and the data used to produce the results presented in this paper can be found at \href{https://github.com/jonasvoeller/sphericalLattices.git}{GitHub}.

%Our results show that there is an interesting connection between lattice geometry, spectral properties, and the couplings of interacting theories.
%The fact that a free-particle spectral optimization leads to geometries satisfying the uniformity conditions required by the isoradial construction suggests that a more general connection between spectral optimization and critical couplings for nearest-neighbor interacting models may exist.

%In the particular case of the TFIM, Ising couplings provide a better approximation to $\mathrm{SO}(3)$ than Laplacian couplings, which indicates that their form closer approximates the correct coupling construction needed.
%As already mentioned, we assume that the correct couplings depend nontrivially on the transverse field, which is not yet known and is therefore a natural direction for future work.
%Instead of adjusting the couplings, we tried using a non-constant transverse field, adjusting $h$ with the Voronoi cell area of each vertex, which however resulted in increased spherical symmetry breaking.

%More generally, it would be interesting to investigate whether analogous optimization strategies can be applied to other interacting lattice models and to spherical geometries in higher dimensions.

\section*{Acknowledgements}
We thank F. Parisen Toldin for  discussions and acknowledge computing time at the NHR Center at RWTH Aachen University.

\begin{appendix}
\numberwithin{equation}{section}

\section{Decomposition of \texorpdfstring{$\mathrm{O}(3)$}{O(3)} to \texorpdfstring{$I_h$}{Ih}}
\label{sec:grouptheorydetails}
The full orthogonal symmetry group of the sphere, including both rotations and inversion, is the infinite group $\mathrm{O}(3)$, whereas the refined base icosahedral lattices considered here retain only the discrete symmetry group $I_h$ of the icosahedron.
Consequently, a continuum multiplet with angular momentum quantum number $\ell$ will generally not remain degenerate on the lattice but split into several multiplets that transform according to irreducible representations of $I_h$.
The decomposition of the $(2\ell+1)$-dimensional irreducible representation of $\mathrm{O}(3)$ into irreducible representations of $I_h$ therefore determines the symmetry structure and possible degeneracies of the corresponding free-particle energy levels on the icosahedral lattices.

The multiplicity $n_{\Gamma}$ of an $I_h$ irrep $\Gamma$ restricted to the $\ell$-th reducible representation of $\mathrm{O}(3)$ is obtained from the character orthogonality relation~\cite{2012_inui}
\begin{equation}\label{eq:decomposition}
    n_{\Gamma}=\frac{1}{\lvert I_h\rvert}\sum_C N_C\chi_\ell(C)\chi_\Gamma^*(C),
\end{equation}
where the sum runs over all conjugacy classes $C$ of $I_h$, $N_C$ denotes the number of elements in the class, and $\chi_\ell$ and $\chi_\Gamma$ are the corresponding characters of the reducible and irreducible representations, respectively.
Furthermore, the relation is normalized by the size of the finite group $\lvert I_h\rvert=120$.

\subsection{Character table of \texorpdfstring{$I_h$}{Ih}}
The full icosahedral group can be written as the direct product
\begin{equation}
	I_h=I\times C_i,
\end{equation}
where $I$ denotes the rotational icosahedral group and $C_i=\{E,i\}$ contains the identity and inversion.
Since inversion commutes with every element of $I$, each irreducible representation of $I_h$ is obtained by assigning a definite parity (even $g$ or odd $u$) to an irreducible representation of $I$.
This doubles the number of irreps while leaving their dimensions unchanged.

The character table for $I_h$ is shown in \cref{tab:Ihcharactertable}, where we indicate the character table for $I$ with a gray background.
The conjugacy classes of $I$ consist of the identity, as well as rotations by $2\pi/5$, $4\pi/5$, $2\pi/3$ and $\pi$, denoted by $C_5$, $C_5^2$, $C_3$ and $C_2$.
Combining these operations with inversion generates the remaining classes of $I_h$, namely the improper rotations $S_{10}, S_{10}^3$ and $S_6$, as well as the reflection $\sigma$.
For both the even and odd sectors, the possible dimensions of irreps are $[1,3,3,4,5]$, which determine the degeneracies that can occur in the energy spectrum once the continuous rotational symmetry is reduced to the discrete icosahedral symmetry.
\begin{table}[]
	\centering
	\begin{tabular}{c|cccccccccc}
		$I_h$ & $E$ & $12C_5$ & $12C_5^2$ & $20C_3$ & $15C_2$ & $i$ & $12S_{10}$ & $12S_{10}^3$ & $20S_6$ & $15\sigma$ \\
		\hline
		$A_g$     & \cellcolor{gray!30} 1 & \cellcolor{gray!30} 1 & \cellcolor{gray!30} 1 & \cellcolor{gray!30} 1 & \cellcolor{gray!30} 1 & 1 & 1 & 1 & 1 & 1 \\
		$T_{1g}$  & \cellcolor{gray!30} 3 & \cellcolor{gray!30} $\phi$ & \cellcolor{gray!30} $\phi'$ & \cellcolor{gray!30} 0 & \cellcolor{gray!30} -1 & 3 & $\phi'$ & $\phi$ & 0 & -1 \\
		$T_{2g}$  & \cellcolor{gray!30} 3 & \cellcolor{gray!30} $\phi'$ & \cellcolor{gray!30} $\phi$ & \cellcolor{gray!30} 0 & \cellcolor{gray!30} -1 & 3 & $\phi$ & $\phi'$ & 0 & -1 \\
		$G_g$     & \cellcolor{gray!30} 4 & \cellcolor{gray!30} -1 & \cellcolor{gray!30} -1 & \cellcolor{gray!30} 1 & \cellcolor{gray!30} 0 & 4 & -1 & -1 & 1 & 0 \\
		$H_g$     & \cellcolor{gray!30} 5 & \cellcolor{gray!30} 0 & \cellcolor{gray!30} 0 & \cellcolor{gray!30} -1 & \cellcolor{gray!30} 1 & 5 & 0 & 0 & -1 & 1 \\
		$A_u$     & 1 & 1 & 1 & 1 & 1 & -1 & -1 & -1 & -1 & -1 \\
		$T_{1u}$  & 3 & $\phi$ & $\phi'$ & 0 & -1 & -3 & $-\phi'$ & $-\phi$ & 0 & 1 \\
		$T_{2u}$  & 3 & $\phi'$ & $\phi$ & 0 & -1 & -3 & $-\phi$ & $-\phi'$ & 0 & 1 \\
		$G_u$     & 4 & -1 & -1 & 1 & 0 & -4 & 1 & 1 & -1 & 0 \\
		$H_u$     & 5 & 0 & 0 & -1 & 1 & -5 & 0 & 0 & 1 & -1 \\
	\end{tabular}
	\caption{Character Table of the Icosahedral Group $I_h$ with the golden ratio $\phi = (1+\sqrt{5})/2$ and $\phi' = 1-\phi = (1-\sqrt{5})/2$.
	The uppermost row shows all ten classes of the group, while the leftmost column shows all irreducible representations.
	Cells with a gray background belong to the icosahedral group without inversion $I$.}\label{tab:Ihcharactertable}
\end{table}

\subsection{Restriction of \texorpdfstring{$\mathrm{O}(3)$}{O(3)} irreps to \texorpdfstring{$I_h$}{Ih}}
We now consider the restriction of the irreducible representations $\Gamma_{\ell}$ of $\mathrm{O}(3)$ to the subgroup $I_h$.
The $\ell$-th irrep acts on a $(2\ell+1)$-dimensional space spanned by the spherical harmonics $Y_{\ell m}$.
Since the identity element acts trivially on a representation in that space, the character for $E$ is simply given by $\chi_{\ell}(E)=2\ell+1$.

For a proper rotation by an angle $\alpha\neq0$, the spherical harmonics transform according to $C_\alpha Y_{\ell m}(\theta, \phi)=e^{-im\alpha}Y_{\ell m}(\theta, \phi)$.
The corresponding character is obtained by summing over all $m$
\begin{equation}\label{eq:characterformular}
	\chi_\ell(\alpha) = \sum_{m=-\ell}^{\ell} e^{-im\alpha} = \frac{\sin\left(\left(\ell + \frac{1}{2}\right)\alpha\right)}{\sin(\alpha/2)}.
\end{equation}

For improper elements, inversion contributes a factor determined by the parity of the spherical harmonics.
Since $Y_{\ell m}(-\mathbf{r})=(-1)^\ell Y_{\ell m}(\mathbf{r})$, an improper element can be written as $X=iR$, where $R\in I$ is the corresponding proper rotation.
Its character is therefore
\begin{equation}
    \chi_\ell(X)=(-1)^\ell\chi_\ell(\alpha_R),
\end{equation}
where $\alpha_R$ is the rotation angle of the proper rotation $R$ in the decomposition $X=iR$.
Thus, for the improper classes in \cref{tab:Ihcharactertable}, the angles used in evaluating the characters are those of their corresponding proper rotations.

\subsection{Decomposition for \texorpdfstring{$\ell=0, 1,\ldots,7$}{ell=0,1,...,7}}
As an explicit example, we consider the restriction of the $\ell=3$ representation of $\mathrm{O}(3)$ to $I_h$ by applying \cref{eq:decomposition}.
Since $\ell=3$ is odd, only irreducible representations with odd parity can appear in the decomposition.
The restricted characters, obtained from \cref{eq:characterformular} and the improper-element rule above, are listed in \cref{tab:charactertablel3}.
Inserting these together with the $I_h$ character \cref{tab:Ihcharactertable} yields the multiplicities
\begin{table}[]
\centering
\begin{tabular}{c|cc}
	$I_h$ & \textbf{Rotational Angle of $R$($\alpha_R$)}& $\chi_{\text{red}}(C)$ \\
	\hline
	$E$ & $0$ & $7$ \\
	$12C_5$ & $2\pi/5$ & $-\phi$ \\
	$12C_5^2$ & $4\pi/5$ & $-\phi'$ \\
	$20C_3$ & $2\pi/3$ & $1$ \\
	$15C_2$ & $\pi$ & $-1$ \\
	$i$ & $0$& $-7$ \\
	$12S_{10}$ & $4\pi/5$ & $\phi'$ \\
	$12S_{10}^3$ & $2\pi/5$& $\phi$ \\
	$20S_6$ & $2\pi/3$& $-1$ \\
	$15\sigma$ & $\pi$& $1$ \\
\end{tabular}
\caption{Characters of the $\ell=3$ $\mathrm{O}(3)$ irrep restricted to $I_h$ operations.}\label{tab:charactertablel3}
\end{table}
\begin{equation}
    \begin{aligned}
    n_{T_{2u}} &= \frac{1}{120}\Big[21-12\phi\phi'-12\phi\phi'+15+21-12\phi\phi'-12\phi\phi'+15\Big] \\
    &= \frac{1}{120}\left(42-48\phi\phi'+30\right) \\
    &= \frac{1}{120}(72+48)=1, \\
    n_{G_u}
    &= \frac{1}{120}\Big[28+12\phi+12\phi'+20+28+12\phi'+12\phi+20\Big] \\
    &= \frac{1}{120}\left(56+24(\phi+\phi')+40\right) \\
    &= \frac{1}{120}(96+24)=1, \\
    n_{A_u}&=n_{T_{1u}}=n_{H_u}=0,
    \end{aligned}
\end{equation}
where we use the identities $\phi+\phi'=1$ and $\phi\phi'=-1$.
Hence, the $\ell=3$ representation decomposes as
\begin{equation}
    \Gamma_{\ell=3}\to T_{2u}\oplus G_u.
\end{equation}

The same procedure can be applied to the remaining values of $\ell$.
The resulting decompositions for $\ell=0, 1, \ldots, 7$ are summarized in \cref{tab:deompositionl0to7}.
\begin{table}[h!]
    \centering
    \begin{tabular}{c|c|l|l}
        $\ell$ & $\dim = 2\ell+1$ & Decomposition under $I_h$ & Dimensions of irreps \\
        \hline
        $0$ & $1$ & $A_g$ & $1$\\
        $1$ & $3$ & $T_{1u}$ & $3$\\
        $2$ & $5$ & $H_g$ & $5$\\
        $3$ & $7$ & $T_{2u} \oplus G_u$ & $3+4$\\
        $4$ & $9$ & $G_g \oplus H_g$ & $4+5$\\
        $5$ & $11$ & $T_{1u} \oplus T_{2u} \oplus H_u$ & $3+3+5$\\
        $6$ & $13$ & $A_g \oplus T_{1g} \oplus G_g \oplus H_g$ & $1+3+4+5$ \\
        $7$ & $15$ & $T_{1u} \oplus T_{2u} \oplus G_u \oplus H_u$ & $3+3+4+5$\\
    \end{tabular}
    \caption{Decomposition of $\ell$-th $\mathrm{O}(3)$ irreps under $I_h$ for $\ell=0, 1,\ldots,7$.}\label{tab:deompositionl0to7}
\end{table}

\section{Calculation of \texorpdfstring{$\nabla_{\mathbf{r}_k}J_{ij}$}{grad J}}
\label{sec:gradientderivation}
In this section, we derive the full analytical gradient of the free-particle hopping coefficients with respect to vertex positions $\mathbf{r}$.

In the following, if $\nabla_{\mathbf{r}_k}$ acts on a scalar, it produces the usual column vector of partial derivatives.
If it instead acts on a vector-valued function $\mathbf{v}(\mathbf{r}_k)\in\mathbb{R}^3$, it produces the $3\times3$ Jacobian matrix $\mathbf{A}$ with entries $A_{pq}=\partial v_p/\partial r_{k,q}$.
We write ${[\mathbf{v}]}_\times$ for the skew-symmetric cross-product matrix satisfying ${[\mathbf{v}]}_\times\mathbf{u} = \mathbf{v}\times\mathbf{u}$.

The hopping coefficients are defined in \cref{eq:laplaciancouplings} as half the ratio of the dual edge length $l^*_{ij}$ to its corresponding primary edge length $l_{ij}$.
Applying the quotient rule gives
\begin{equation}\label{eq:gradientoff-diagonal}
	\nabla_{\mathbf{r}_k} J_{ij}
    = \frac{l_{ij}\,\nabla_{\mathbf{r}_k}l_{ij}^*
           - l_{ij}^*\,\nabla_{\mathbf{r}_k}l_{ij}}
          {2\,l_{ij}^2}.
\end{equation}
We now compute each gradient on the right-hand side of \cref{eq:gradientoff-diagonal}.
Constructing the full gradient is then simply a task of assembling each contribution in the end.

\subsection{Gradient of \texorpdfstring{$l_{ij}$}{l}}
This gradient is straightforward and takes the following form:
\begin{equation}
    \nabla_{\mathbf{r}_k}l_{ij} =
    \begin{cases}
        \dfrac{\mathbf{r}_i - \mathbf{r}_j}{\lvert \mathbf{r}_i - \mathbf{r}_j \rvert},
            & k = i, \\[1em]
        -\dfrac{\mathbf{r}_i - \mathbf{r}_j}{\lvert \mathbf{r}_i - \mathbf{r}_j \rvert},
            & k = j, \\[1em]
        \mathbf{0}, & \text{otherwise}.
    \end{cases}
\end{equation}

\subsection{Gradient of \texorpdfstring{$l^*_{ij}$}{l*}}
This gradient is more complex than the previous one.
In the framework of Regge calculus, which we assume in \cref{sec:latticeoptimization}, the distance $l^*_{ij}$ on the dual lattice is the sum of the path lengths from the circumcenter of each contributing triangle to the midpoint between vertices $i$ and $j$ on the triangular lattice.
These distances are calculated on the faces, not as geodesics.
A full definition is given by
\begin{equation}
	l_{ij}^*(\mathbf{r}_i, \mathbf{r}_j, \mathbf{r}_l, \mathbf{r}_m) = |\mathbf{p}_{ij}(\mathbf{r}_i, \mathbf{r}_j)-\mathbf{C}_{ijl}(\mathbf{r}_i, \mathbf{r}_j, \mathbf{r}_l)|+|\mathbf{p}_{ij}(\mathbf{r}_i, \mathbf{r}_j)-\mathbf{C}_{ijm}(\mathbf{r}_i, \mathbf{r}_j, \mathbf{r}_m)|,
\end{equation}
where $\mathbf{p}_{ij}=(\mathbf{r}_i+\mathbf{r}_j)/2$ is the midpoint between the vertices $i$ and $j$, and $\mathbf{C}_{ijl}$ and $\mathbf{C}_{ijm}$ are the circumcenters of the triangles defined by the points $\mathbf{r}_i, \mathbf{r}_j, \mathbf{r}_l$ and $\mathbf{r}_i, \mathbf{r}_j, \mathbf{r}_m$, respectively.
Differentiating gives
\begin{equation}
	\nabla_{\mathbf{r}_k}l_{ij}^* = (\nabla_{\mathbf{r}_k}(\mathbf{p}_{ij}-\mathbf{C}_{ijl}))^\top\frac{\mathbf{p}_{ij}-\mathbf{C}_{ijl}}{|\mathbf{p}_{ij}-\mathbf{C}_{ijl}|}
    +(\nabla_{\mathbf{r}_k}(\mathbf{p}_{ij}-\mathbf{C}_{ijm}))^\top\frac{\mathbf{p}_{ij}-\mathbf{C}_{ijm}}{|\mathbf{p}_{ij}-\mathbf{C}_{ijm}|},
\end{equation}
where the two terms are structurally identical, so we work out only the first.
The Jacobian of the midpoint is
\begin{equation}
	\nabla_{\mathbf{r}_k}\mathbf{p}_{ij} =
	\begin{cases}
		\frac{1}{2}\mathbf{I}, & k = i,j, \\
		\mathbf 0, & \text{otherwise}.
	\end{cases}
\end{equation}
What remains is the computation of the Jacobian of $\mathbf{C}_{ijl}$.

\subsection{Gradient of the circumcenter \texorpdfstring{$\mathbf{C}_{ijl}$}{C}}
In three dimensions, the circumcenter of a triangle is given by the following formula~\cite{EppsteinCircumcenter}:
\begin{equation}
	\mathbf{C}_{ijl}(\mathbf{r}_i, \mathbf{r}_j, \mathbf{r}_l) = \mathbf{r}_i + \frac{|\mathbf{r}_{il}|^2(\mathbf{n}\times\mathbf{r}_{ij})-|\mathbf{r}_{ij}|^2(\mathbf{n}\times\mathbf{r}_{il})}{2|\mathbf{n}|^2},
\end{equation}
with $\mathbf{r}_{i\alpha}=\mathbf{r}_{\alpha}-\mathbf{r}_{i}$, where $\alpha=j,l$, and $\mathbf{n}=\mathbf{r}_{ij}\times\mathbf{r}_{il}$.
To keep the notation compact we define
\begin{equation}
    \boldsymbol{\xi}
    \;=\;
    |\mathbf{r}_{il}|^2(\mathbf{n}\times\mathbf{r}_{ij})
    - |\mathbf{r}_{ij}|^2(\mathbf{n}\times\mathbf{r}_{il}),
    \quad \text{and} \quad
    d = 2|\mathbf{n}|^2,
\end{equation}
so that $\mathbf{C}_{ijl} = \mathbf{r}_i + \boldsymbol{\xi}/d$.
The quotient rule then gives
\begin{equation}\label{Eq: gradC}
    \nabla_{\mathbf{r}_k}\mathbf{C}_{ijl}
    = \delta_{ki}\,\mathbf{I}
    + \frac{1}{d}\,\nabla_{\mathbf{r}_k}\boldsymbol{\xi}
    - \frac{\boldsymbol{\xi}}{d^2}\,{(\nabla_{\mathbf{r}_k}d)}^\top,
\end{equation}
where $\boldsymbol{\xi}/d^2$ times ${(\nabla_{\mathbf{r}_k}d)}^\top$ is an outer product
(a $3\times 3$ rank-1 matrix).
At this point, there are two remaining gradients that need to be calculated.
The second one, $\nabla_{\mathbf{r}_k}d$, is rather simple and takes the form
\begin{equation}\label{eq:grad_d}
    \nabla_{\mathbf{r}_k}d
    = 4\,{(\nabla_{\mathbf{r}_k}\mathbf{n})}^{\top}\mathbf{n},
\end{equation}
where $\nabla_{\mathbf{r}_k}\mathbf{n}$ is the $3\times 3$ Jacobian
\begin{equation}
    \nabla_{\mathbf{r}_k}\mathbf{n} =
    \begin{cases}
        {[\mathbf{r}_{il}]}_\times - {[\mathbf{r}_{ij}]}_\times,& k = i, \\
        -{[\mathbf{r}_{il}]}_\times,                            & k = j, \\
        {[\mathbf{r}_{ij}]}_\times,                 			& k = l, \\
        \mathbf{0},                                           	& \text{otherwise}.
    \end{cases}
\end{equation}
Lastly, the gradient of $\boldsymbol{\xi}$ works out to 
\begin{equation}
	\begin{aligned}
		\nabla_{\mathbf{r}_k}\boldsymbol{\xi}
		&=
		2{(\mathbf{n} \times \mathbf{r}_{ij})}\left[{(\nabla_{\mathbf{r}_k}\mathbf{r}_{il})}^\top \mathbf{r}_{il}\right]^\top
		+ |\mathbf{r}_{il}|^2
		\left(
		- {[\mathbf{r}_{ij}]}_\times \nabla_{\mathbf{r}_k}\mathbf{n}
		+ {[\mathbf{n}]}_\times \nabla_{\mathbf{r}_k}\mathbf{r}_{ij}
		\right) \\
		&+2{(\mathbf{r}_{il} \times \mathbf{n})}\left[(\nabla_{\mathbf{r}_k}\mathbf{r}_{ij})^\top \mathbf{r}_{ij}\right]^\top
		 + |\mathbf{r}_{ij}|^2
		\left(
		{[\mathbf{r}_{il}]}_\times \nabla_{\mathbf{r}_k}\mathbf{n}
		- {[\mathbf{n}]}_\times \nabla_{\mathbf{r}_k}\mathbf{r}_{il}
		\right),
	\end{aligned}
\end{equation}
where
\begin{equation*}
	\nabla_{\mathbf{r}_k}\mathbf{r}_{ij}
	=
	\begin{cases}
		-\mathbf{I}, & k = i, \\
		+\mathbf{I}, & k = j, \\
		\mathbf{0}, & \text{otherwise},
	\end{cases}
	\quad \text{and} \quad	
	\nabla_{\mathbf{r}_k}\mathbf{r}_{il}
	=
	\begin{cases}
		-\mathbf{I}, & k = i, \\
		+\mathbf{I}, & k = l, \\
		\mathbf{0}, & \text{otherwise}.
	\end{cases}
\end{equation*}

\subsection{\texorpdfstring{$\mathcal{O}(N)$}{O(N)} scaling}
\begin{figure}[h!]
	\centering
	\begin{tikzpicture} [scale=1.0]
		\def\a{1.5} % Lattice spacing

		% Central point
		\coordinate (C) at (0,0);

		% Define and draw the triangular grid
		\foreach \i in {1,...,6} {
			\coordinate (P\i) at ({2*cos(60*\i)}, {2*sin(60*\i)});
		\foreach \j in {-2, 0, 2} {
			\coordinate (MiddleRow\i\j) at ({2*cos(60*\i)-\j}, {2*sin(60*\i)});
		}
		\foreach \j in {-3, -1, 1, 3} {
			\coordinate (BottomRow\i\j) at ({2*cos(60*\i)-\j}, {2*sin(60*\i)-sqrt(3)});
			\coordinate (BottomRowC\i\j) at ({-\j}, {-sqrt(3)});
			\coordinate (TopRow\i\j) at ({2*cos(60*\i)-\j}, {2*sin(60*\i)+sqrt(3)});
			\coordinate (TopRowC\i\j) at ({-\j}, {sqrt(3)});
		}
		}

		\foreach \i/\j in {1/2, 2/3, 3/4, 4/5, 5/6, 6/1} {
		\foreach \k in {-2, 0, 2} {
			\draw[gray!50, thin] (MiddleRow\i\k) -- (MiddleRow\j\k);
		}
		\foreach \k in {-3, -1, 1, 3} {
			\draw[gray!50, thin] (BottomRow\i\k) -- (BottomRow\j\k);
			\draw[gray!50, thin] (BottomRow\i\k) -- (BottomRowC\i\k);
			\draw[gray!50, thin] (TopRow\i\k) -- (TopRow\j\k);
			\draw[gray!50, thin] (TopRow\i\k) -- (TopRowC\i\k);
		}
		}

		% Define and draw the important circumcenters
		\foreach \i in {1,...,6} {
			\coordinate (InnerCircumcenter\i) at ({2*0.55*cos(60*\i+30)}, {2*0.55*sin(60*\i+30)});
			\coordinate (OuterCircumcenter\i) at ({2*1.2*cos(60*\i+30)}, {2*1.2*sin(60*\i+30)});
			\filldraw[solid, color=gray!80, fill=gray!0, thin] (InnerCircumcenter\i) circle(3pt);
			\filldraw[solid, color=gray!80, fill=gray!0, thin] (OuterCircumcenter\i) circle(3pt);
		}

		% Draw the inner hij
		\foreach \i in {1,...,6} {
			\draw[solid, blue, semithick, double] (C) -- (P\i);
		}

		% Draw the outer hij and inner sij
		\foreach \i/\j in {1/2, 2/3, 3/4, 4/5, 5/6, 6/1} {
			\draw[solid, Red, semithick, double] (P\i) -- (P\j);
			\draw[densely dashed, ForestGreen, semithick, double] (InnerCircumcenter\i) -- (InnerCircumcenter\j);
		}

		% Draw the outer sij
		\foreach \i in {1,...,6} {
			\draw[densely dashed, YellowOrange, semithick, double] (OuterCircumcenter\i) -- (InnerCircumcenter\i);
		}

		% Define and draw the triangular grid
		\foreach \i in {1,...,6} {
		\foreach \j in {-2, 0, 2} {
			\filldraw[black!20] (MiddleRow\i\j) circle (1.6pt);
		}
		\foreach \j in {-3, -1, 1, 3} {
			\filldraw[black!20] (BottomRow\i\j) circle (1.6pt);
			\filldraw[black!20] (TopRow\i\j) circle (1.6pt);
		}
		};
        \filldraw[black!80] (MiddleRow10) circle (2.2pt);
        \filldraw[black!80] (MiddleRow20) circle (2.2pt);
        \filldraw[black!80] (MiddleRow30) circle (2.2pt);
        \filldraw[black!80] (MiddleRow40) circle (2.2pt);
        \filldraw[black!80] (MiddleRow50) circle (2.2pt);
        \filldraw[black!80] (MiddleRow60) circle (2.2pt);
        
        \filldraw[black!80] (MiddleRow42) circle (2.2pt);
        \filldraw[black!80] (MiddleRow22) circle (2.2pt);
        \filldraw[black!80] (MiddleRow5-2) circle (2.2pt);
        \filldraw[black!80] (MiddleRow1-2) circle (2.2pt);

        \filldraw[black!80] (TopRow11) circle (2.2pt);
        \filldraw[black!80] (BottomRow51) circle (2.2pt);
        
		\filldraw[black] (C) circle (3.5pt) node[anchor=north west]{$\mathbf{r}_k$};
	\end{tikzpicture}
	\caption[Vertex connections involved in the calculation of $\nabla_{\mathbf{r}_k}J_{ij}$ for the loss function used in the gradient descent]{Regular triangular lattice to serve as a sketch of a part of a triangulated spherical lattice.
    In the center, we show the vertex $k$ at position $\mathbf{r}_k$, which has coordination number $n_c=6$.
    All $2n_c=12$ vertices that are in connection with hopping coefficients whose values depend on $\mathbf{r}_k$ are shown as black dots.
    The circumcenters of the corresponding triangles are shown as open circles.
	The blue double lines connecting $\mathbf{r}_k$ with its nearest neighbors denote all $l_{ij}$ that depend on $\mathbf{r}_k$.
	The red double lines show all $l_{ij}$ that do not depend on $\mathbf{r}_k$, but are part of a hopping element $J_{ij}$ that involves $\mathbf{r}_k$ via the circumcenter of a triangle that shares that specific edge $\langle ij\rangle$.
	The green dashed double lines show all $l^*_{ij}$ of hopping elements that involve the edge length $l_{ij}$ shown in blue, and the orange dashed double lines show all $l^*_{ij}$ of hopping elements that involve the edge length $l_{ij}$ shown in red.}\label{Fig: neighbors}
\end{figure}
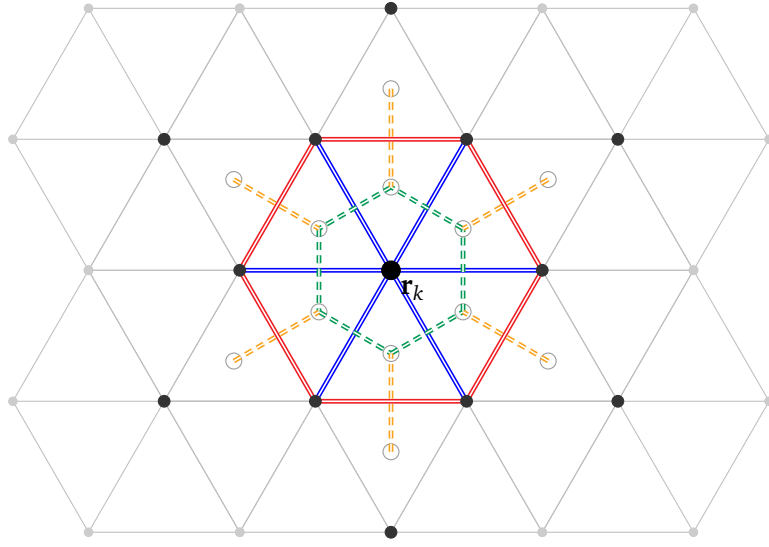

Collecting all cases that give a contribution, the gradient $\nabla_{\mathbf{r}_k} J_{ij}$ is non-zero for three types of vertices: the vertex $k$ itself, the direct neighbors of that vertex, and all vertices that are part of a triangle whose circumcenter enters into the calculation of the gradient.
This is illustrated in \cref{Fig: neighbors}.
Since each vertex has either $5$ or $6$ neighbors, its gradient will depend on either $10$ or $12$ vertices in total.
With that, the complexity of the total gradient scales as $\mathcal{O}(N)$.

\section{Details on the gradient descent optimization}
\label{sec:laplacianappendix}
To optimize the vertex positions of our lattices, we minimize the loss function in \cref{eq:lossfunction} via simple gradient descent.
Each gradient descent step is performed using a fixed learning rate and is followed by a radial projection of all vertices back onto the unit sphere, thereby preserving the spherical constraint throughout the optimization.
We verified that the optimization preserves the initial triangulation, so only the vertex positions evolve.

To perform the optimization, it is important to choose a suitable number of energy levels to optimize with respect to.
During the optimization process, we observe that higher energy levels have a higher sensitivity to small vertex displacements.
Including too few energy levels leaves room for further optimization, while including too many energy levels leads to a breakdown of the process.
In practice, we optimize approximately $\sqrt{N}/2$ multiplets for lattices with $N<1000$.
For larger lattices, we found that fewer levels result in a more uniform geometry and a reliable improvement of the low-lying energy spectrum.
It therefore proves useful to start the optimization with a conservative number of energy multiplets (we used twelve).
After an initial optimization, the number of multiplets can then be increased to get an improved result.
(The code we used and the resulting spherical lattices can be found at \href{https://github.com/jonasvoeller/sphericalLattices.git}{GitHub}.)

\begin{figure}[h!]
    \centering
    \begin{subfigure}[t]{0.49\textwidth}
        \centering
        \includegraphics[width=0.95\textwidth]{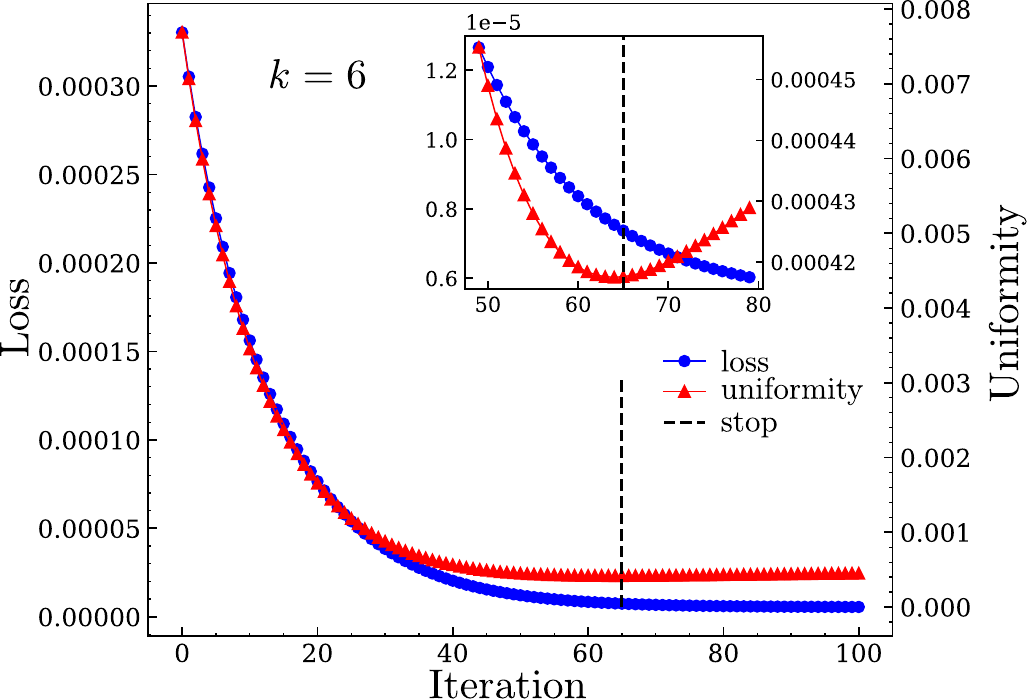}
    \end{subfigure}
    \begin{subfigure}[t]{0.49\textwidth}
        \centering
        \includegraphics[width=0.95\textwidth]{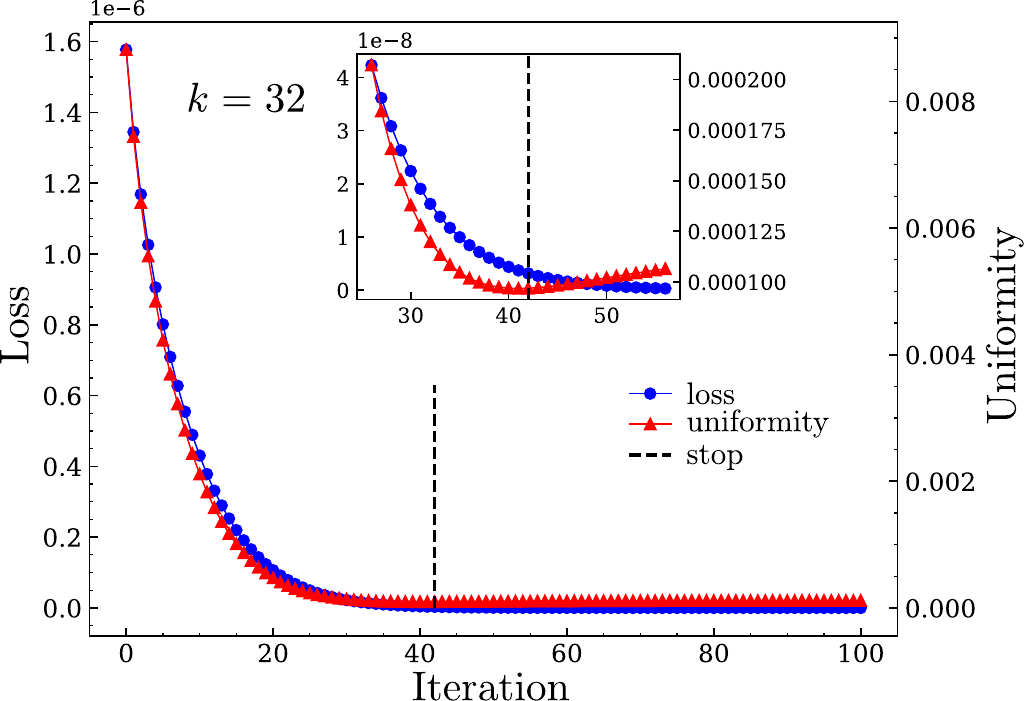}
    \end{subfigure}
    \caption{Loss (blue dots) and uniformity (red triangles) during optimization for lattices with level of subdivision $k=6$ (left) and $k=32$ (right).
    The stopping point is indicated as a black dashed line.}\label{fig:loss}
\end{figure}

Moreover, we argued in \cref{sec:geometriceffects} that the uniformity in circumradii and perimeters improves together with the spectral loss up to a certain point which we defined as our stopping criterion.
This is shown in \cref{fig:loss}, where we define $\text{Uniformity}=\nu_R+\nu_P$.
Beyond the stopping point, the spectral loss continues to decrease only marginally, while the geometric uniformity deteriorates slightly.
This produces lattices that retain nearly optimal spectral properties while simultaneously satisfying the geometric requirements for the interacting models studied in \cref{sec:isingmodelons2} and \cref{sec:pottsmodelons2}.

\section{Robustness of the Ising rotational symmetry breaking measurements}
\label{sec:isingappendix}
In this section, we go into further detail on the rotational symmetry breaking measurements performed in \cref{sec:isingmodelons2}, exploring rotational symmetry in closely related systems.
In particular, we examine the dependence of the symmetry-breaking observable on the orientation of the lattice, the choice of couplings, and the distance from the critical point.
Together, these measurements clarify which ingredients are necessary for recovering the continuum $\mathrm{SO}(3)$ symmetry.

\subsection{Dependence on lattice orientation}
The observables $C_{\ell m}$ defined in \cref{eq:rotsymmobservable} depend on the choice of coordinate system through the spherical harmonics $Y_{\ell m}(\hat{\mathbf{n}}_i)$.
As mentioned in \cref{sec:isingmodelons2}, we align the $z$-axis with a fivefold symmetry axis of the original icosahedron, corresponding to a pair of opposite vertices.
This choice minimizes the mixing between irreducible representations of the icosahedral group $I_h$ and therefore provides the clearest separation between symmetry-protected and symmetry-broken sectors.

\begin{figure}[h!]
    \centering
    \begin{subfigure}[t]{0.49\textwidth}
        \centering
        \includegraphics[width=0.95\textwidth]{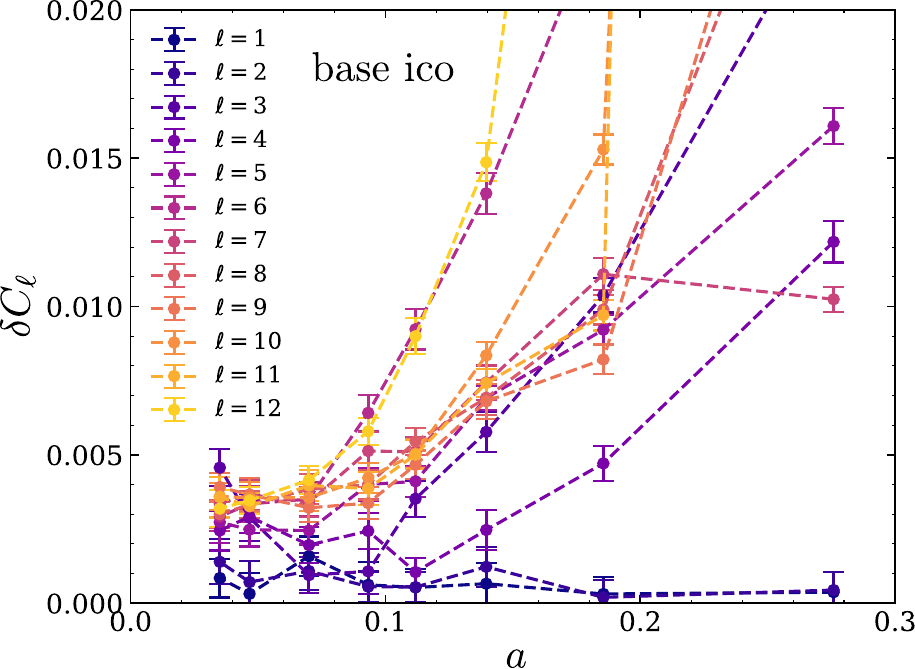}
    \end{subfigure}
    \begin{subfigure}[t]{0.49\textwidth}
        \centering
        \includegraphics[width=0.95\textwidth]{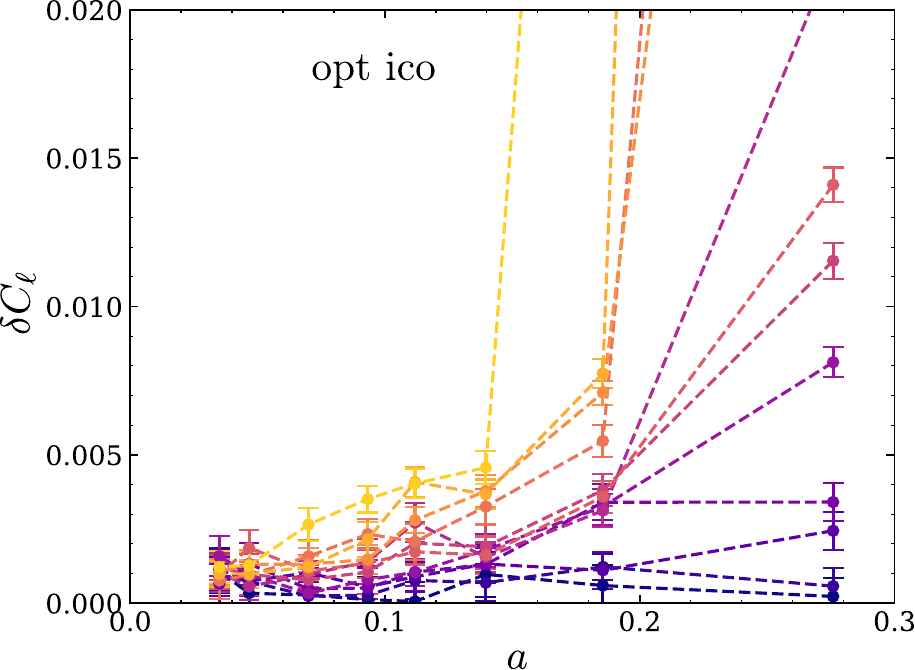}
    \end{subfigure}
    \caption{Measurement of the rotational symmetry breaking using the base (left) and the optimized (right) icosahedral lattices with no particular alignment.}\label{fig:symmetrybreakingisingnonaligned}
\end{figure}

Figure~\ref{fig:symmetrybreakingisingnonaligned} shows the same analysis for a generic orientation in which the $z$-axis is not aligned with any symmetry axis of the underlying icosahedron.
While the overall magnitude of $\delta C_\ell$ is slightly reduced compared to the aligned case, the qualitative behavior remains unchanged.
The optimized lattices again exhibit a vanishing symmetry-breaking signal in the continuum limit, whereas the base lattices do not.
The reduced magnitude appears to be a consequence of the increased mixing between different irreducible representations when the coordinate axes are not aligned with a symmetry axis of the lattice.

\subsection{Dependence on the coupling definition}
We next investigate whether optimizing the lattice geometry alone is sufficient to restore rotational symmetry.
To this end, we repeat the measurements using the optimized lattices, but replace the critical Ising couplings by constant nearest-neighbor couplings and by the Laplacian couplings introduced in \cref{eq:laplaciancouplings}.
For both coupling constructions, we tune the model to criticality before measuring the rotational symmetry breaking.

\begin{figure}[h!]
    \centering
    \begin{subfigure}[t]{0.49\textwidth}
        \centering
        \includegraphics[width=0.95\textwidth]{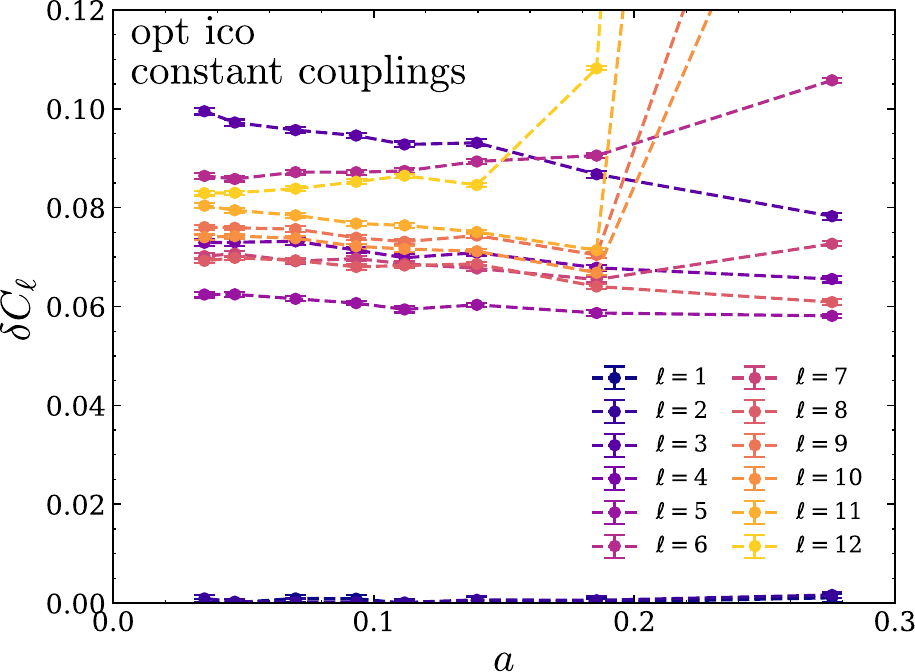}
    \end{subfigure}
    \begin{subfigure}[t]{0.49\textwidth}
        \centering
        \includegraphics[width=0.95\textwidth]{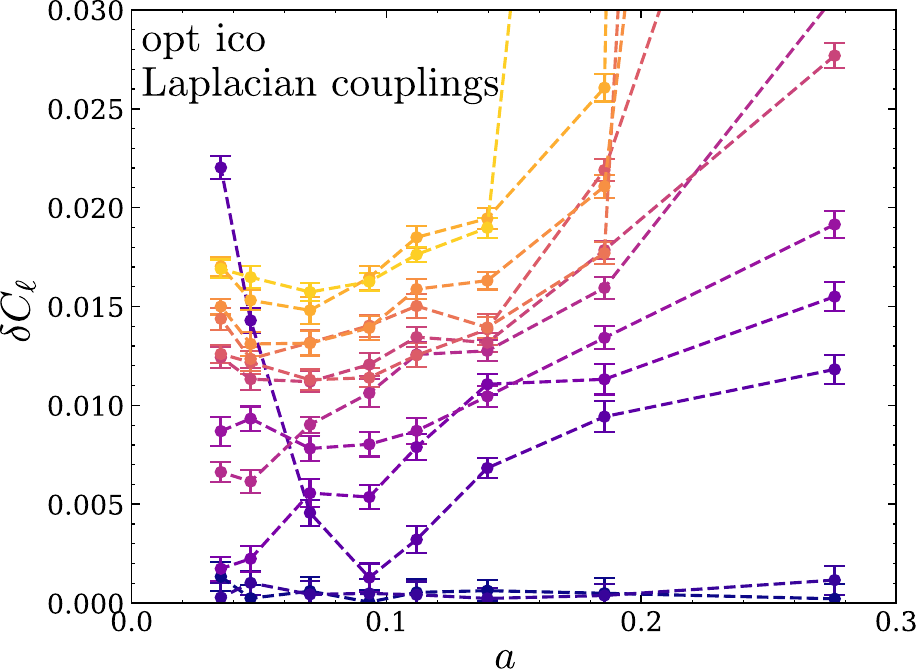}
    \end{subfigure}
    \caption{Measurement of the rotational symmetry breaking using constant (left) and Laplacian (right) couplings combined with the optimized icosahedral lattices.}\label{fig:symmetrybreakingisingdifferentcouplings}
\end{figure}

The resulting measurements are shown in \cref{fig:symmetrybreakingisingdifferentcouplings}.
In both cases, the symmetry breaking observable remains finite in the continuum limit, demonstrating that the optimized lattice geometry alone is insufficient to recover the continuum rotational symmetry.
This is unsurprising for the constant couplings, which do not depend on the lattice geometry and therefore cannot benefit from the optimization at all.
The Laplacian couplings, in contrast, result in a closer approximation to rotational symmetry, which is also not entirely unexpected.
On an affine triangular lattice, the critical Ising couplings satisfy $\sinh(2K_{ij})=l^*_{ij}/l_{ij}$~\cite{2023_Brower_affine}, a relation closely resembling the Laplacian weights.
Nevertheless, this similarity is insufficient to recover the correct continuum behavior.
Neither coupling construction reproduces the behavior obtained with the critical Ising couplings, showing that restoring rotational symmetry requires not only an optimized lattice geometry but also couplings that correctly realize the critical continuum theory.

\subsection{Dependence on temperature}
\begin{figure}[h!]
    \centering
    \begin{subfigure}[t]{0.49\textwidth}
        \centering
        \includegraphics[width=0.95\textwidth]{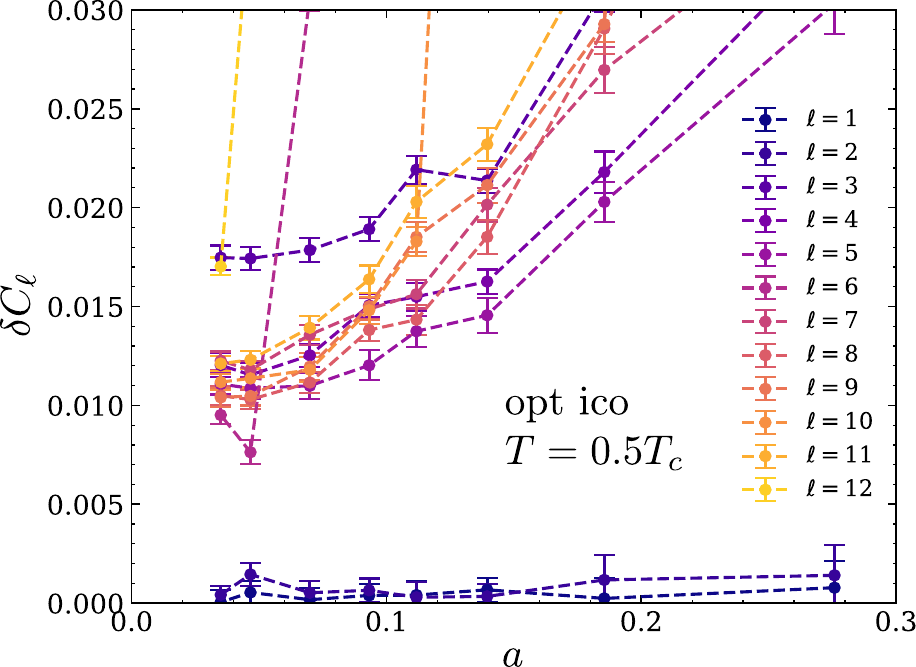}
    \end{subfigure}
    \begin{subfigure}[t]{0.49\textwidth}
        \centering
        \includegraphics[width=0.95\textwidth]{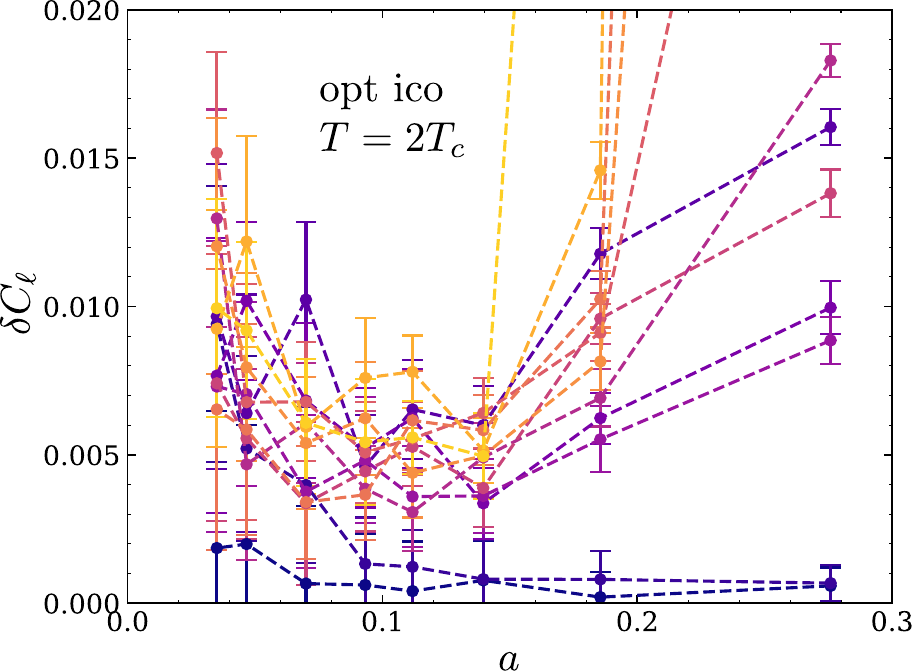}
    \end{subfigure}
    \caption{Measurement of the rotational symmetry breaking using the optimized icosahedral lattices and Ising couplings below the critical point at $T=0.5T_c$ (left) and above the critical point at $T=2T_c$ (right).}\label{fig:symmetrybreakingisingdifferenttemperatures}
\end{figure}

Finally, we investigate whether the restored rotational symmetry persists away from the critical point.
\cref{fig:symmetrybreakingisingdifferenttemperatures} shows the symmetry-breaking observable measured using the optimized lattices with Ising couplings at temperatures below and above criticality.

In both cases, the symmetry-breaking remains finite in the continuum limit.
Thus, despite the optimized lattice geometry and a suitable coupling construction, the system does not recover the full rotational symmetry away from the critical point.
The restoration observed in the main text is therefore a property of the critical fixed point rather than of the lattice geometry alone.

We additionally repeated the measurement above and below the critical point using Laplacian couplings instead of Ising couplings, which however did not improve the results.

Taken together, these additional checks clarify the origin of the restored rotational symmetry observed in the main text.
The optimized lattice geometry is necessary, but it is not sufficient by itself.
Continuum rotational symmetry emerges only when the optimized geometry is combined with couplings that correctly realize the critical fixed point.
Away from criticality, or with inappropriate couplings, significant symmetry breaking remains even in the continuum limit.

%\section{About references}
%Your references should start with the comma-separated author list (initials + last name), the publication title in italics, the journal reference with volume in bold, start page number, publication year in parenthesis, completed by the DOI link (linking must be implemented before publication). If using BiBTeX, please use the style files provided  on \url{https://scipost.org/submissions/author_guidelines}. If you are using our LaTeX template, simply add
%\begin{verbatim}
%\bibliography{your_bibtex_file}
%\end{verbatim}
%at the end of your document. If you are not using our LaTeX template, please still use our bibstyle as
%\begin{verbatim}
%\bibliographystyle{SciPost_bibstyle}
%\end{verbatim}
%in order to simplify the production of your paper.
\end{appendix}


% Provide your bibliography here. You have two options:

%%% FIRST OPTION
% Write your entries here directly, following the example below, including:
% Author(s), Title, Journal Ref. with year in parentheses at the end, followed by the DOI number.

%\begin{thebibliography}{99}
%\bibitem{1931_Bethe_ZP_71} H. A. Bethe, {\it Zur Theorie der Metalle. i. Eigenwerte und Eigenfunktionen der linearen Atomkette}, Zeit. f{\"u}r Phys. {\bf 71}, 205 (1931), \doi{10.1007\%2FBF01341708}.
%\bibitem{arXiv:1108.2700} P. Ginsparg, {\it It was twenty years ago today... }, \url{http://arxiv.org/abs/1108.2700}.
%\end{thebibliography}

% Use your bibtex library, formatted by the SciPost style file.
\bibliography{SciPost_Example_BiBTeX_File.bib}

%%%%%%%%%% END TODO: BIBLIOGRAPHY

\end{document}